\documentclass[journal]{IEEEtran}

\usepackage[cmex10]{amsmath}
\usepackage{amssymb}

\usepackage[pdftex]{graphicx}
\graphicspath{{pics/}}

\ifCLASSINFOpdf
\else
\fi
\begin{document}
%
\title{Tracking of Wideband Multipath Components \\ in a Vehicular Communication Scenario}
%
%
%

\author{Kim~Mahler,~Wilhelm~Keusgen,~Fredrik~Tufvesson,~Thomas~Zemen~and~Giuseppe~Caire
\thanks{Copyright \copyright~2015 IEEE. Personal use of this material is permitted. However, permission to use this material for any other purposes must be obtained from the IEEE by sending a request to pubs-permissions@ieee.org. }
\thanks{K. Mahler and W. Keusgen are with the Department of Wireless Communications and Networks, Fraunhofer Heinrich Hertz Institute, Berlin, Germany. e-mail: (kim.mahler@hhi.fraunhofer.de)}
\thanks{F. Tufvesson is with the Department of Electrical and Information Technology, Lund University, Lund, Sweden.}
\thanks{T. Zemen is with the Digital Safety and Security Department, AIT Austrian Institute of Technology, Vienna, Austria.}
\thanks{G. Caire is with the Communications and Information Theory Group, Technische Universit\"at Berlin, Berlin, Germany.}
\thanks{Manuscript received September 1, 2015.}}

\maketitle

\begin{abstract}
A detailed understanding of the dynamic processes of vehicular radio channels is crucial for its realistic modeling. In this paper, we present multipath components (MPCs) tracking results from a channel sounder measurement with 1 GHz bandwidth at a carrier frequency of 5.7 GHz. We describe in detail the applied algorithms and perform a tracking performance evaluation based on artificial channels and on measurement data from a tunnel scenario. The tracking performance of the proposed algorithm is comparable to the tracking performance of the state-of-the-art Gaussian mixture probability hypothesis density filter, yet with a significantly lower complexity. The fluctuation of the measured channel gain is followed very well by the proposed tracking algorithm, with a power loss of only 2.5 dB. We present statistical distributions for the number of MPCs and the birth/death rate. The applied algorithms and tracking results can be used to enhance the development of geometry-based channel models.
\end{abstract}

%

\IEEEpeerreviewmaketitle
\section{Introduction}
%
%
%
%
\IEEEPARstart{I}{nter-vehicular} radio communication will play an important role in future collision avoidance systems and other intelligent transport system applications, because of its unique potential as a vehicular sensor. Current vehicular sensors, such as radar or video sensors already enable on-board sensor fusion systems to establish a comprehensive perception of the surrounding. However, radio communication enables vehicles to exchange information, even without a direct line-of-sight between the vehicles. 

In order to enhance vehicular communications, researchers focus on a better understanding of the communication performance and its underlying radio propagation processes. Several wideband channel measurements in the 6 GHz band have been conducted, with measurement bandwidths of 60-240 MHz \cite{renaudin08} \cite{karedal10}. In order to extract the relevant channel parameters \cite{bernado14}, complex algorithms are needed \cite{richter05}. However, the complexity of the extraction algorithms can be reduced and the quality of the measurement results enhanced by an increased measurement bandwidth. A sufficiently large measurement bandwidth leads to a sparse multipath channel, where the multipath component (MPC) arrival times can be observed much more accurately than in narrowband channels \cite{molisch05}. Furthermore, due to the fine time resolution, the number of physical MPCs that make up a resolvable MPC is much smaller and results in the absence of significant multipath fading \cite{win98}. Channel sounder measurements around 6 GHz can be referred to as (ultra-)wideband, if the absolute bandwidth exceeds 500 MHz \cite{molisch05}. Channel measurements with a very large bandwidth are today mostly conducted for millimeter wave communications; only little work has been done for vehicle-to-vehicle (V2V) scenarios \cite{lee09}. 

A main benefit of wideband channel sounding measurements is the ability to detect individual MPCs and relate these to physical scattering objects. Consequently, tracking of highly resolved MPCs can be beneficial for a better geometrical understanding of the propagation process in V2V channels. A distinctive characteristic of V2V communication channels is their time-variant behavior, due to the movement of transmitter, receiver and scattering objects. These dynamics lead to a smaller stationarity region of the channel statistics \cite{bernado14} and have to be incorporated appropriately into V2V channel models. Work in \cite{he15} proposes a dynamic V2V channel model based on a local wide-sense stationary (WSS) time window and MPC statistics related to this time window. Alternatively, geometry-based stochastic channel models were proposed, which are well-suited for non-stationary environments \cite{karedal09}. In order to identify the time-variant stochastics of these V2V channel models, researchers focus on tracking the temporal behavior of individual MPCs. 
Work in \cite{czink06} presents an algorithm to track scatterer clusters’ centroids, where the path delay, the angles-of-arrival and the angles-of-departures are used to determine the MPC distance (MCD) in two consecutive time instances, a measure which was first introduced in \cite{soh02}.  The MCD is also used in \cite{he15} as a method to quantify the distance and track MPCs over time. Authors in \cite {froehle12} apply probability hypothesis density (PHD) filters to track MPCs in an indoor UWB channel and claim good tracking performance despite a high amount of diffuse MPCs.

The contribution of this paper is a demonstration that MPCs from wideband V2V channel measurement with a bandwidth of 1 GHz can be tracked with highly accurate estimates and low tracking losses. The proposed tracking algorithm has a low complexity and yet a tracking performance comparable to a state-of-the-art tracking algorithm. We track both the small-scale MPC variations within the time window of an IEEE 802.11p transmission frame and also the behavior of MPCs over longer non-stationary regions. A better understanding of the temporal behavior of individual MPCs will lead to a better geometrical understanding and finally a better representation of the V2V radio propagation process within numerical models. 

This paper is organized as follows: In Section II we introduce our measurement device, the measurement settings and the selected measurement environment. The applied MPC estimation and small-scale tracking algorithm is explained in detail in Section III. Then, we evaluate the performance of these algorithms based on an artificial channel and on empiric measurement data in Section IV. Section V describes a large-scale tracking algorithm required to track MPC across stationarity regions. Finally, we show results of the MPC long-term evolution in a tunnel scenario in Section VI, including statistical distributions on the number of MPCs and their birth/death rate.

\section{Measurement}
\subsection{Measurement Equipment}
The HHI channel sounder, developed at the Fraunhofer Heinrich Hertz Institute (HHI), is a wideband measurement device with a bandwidth of 1 GHz at a carrier frequency of 5.7 GHz \cite{paschalidis08}. The measurement bandwidth allows a delay time resolution of 1 ns (30 cm of wave propagation) and therefore a highly resolved view into the behavior of MPCs. The channel sounder consists of a transmitter unit and a receiver unit that can be installed in conventional passenger vehicles and deployed in real traffic scenarios. For our measurements we use an Audi Avant as transmitter and a Renault Scenic as receiver vehicle. Both vehicles are equipped with omnidirectional and vertically polarized antennas mounted on the roof at the left edge of the vehicle. The deployed antennas were developed at HHI and are matched to the desired frequency range from 5.2 to 6.2 GHz. The antennas feature under laboratory conditions an evenly distributed azimuth radiation pattern with maximum deviations of 0.4 dB from the maximum value and a mean deviation of around 0.2 dB. This radiation pattern probably differs when the antennas are mounted on the roof of the vehicles. However, we expect no significant deviation from the laboratory measurement, due to the fact that the total length of the antenna is around 30 cm and therefore multiple wavelengths above the metallic roof surface. 
In order to record the position of the vehicles during measurements, the highly accurate positioning system GeneSys ADMA-G was used, which works for a limited time even in tunnels without coverage of GPS satellites. Based on measurements in several tunnels with lengths of up to 2 km, we estimate an accuracy of 20 m for the first 1000 m traveled, which applies for the measurement run described in this paper.
\subsection{Measurement Run }
\begin{figure}[!t]
\centering
\includegraphics[width=3.3in]{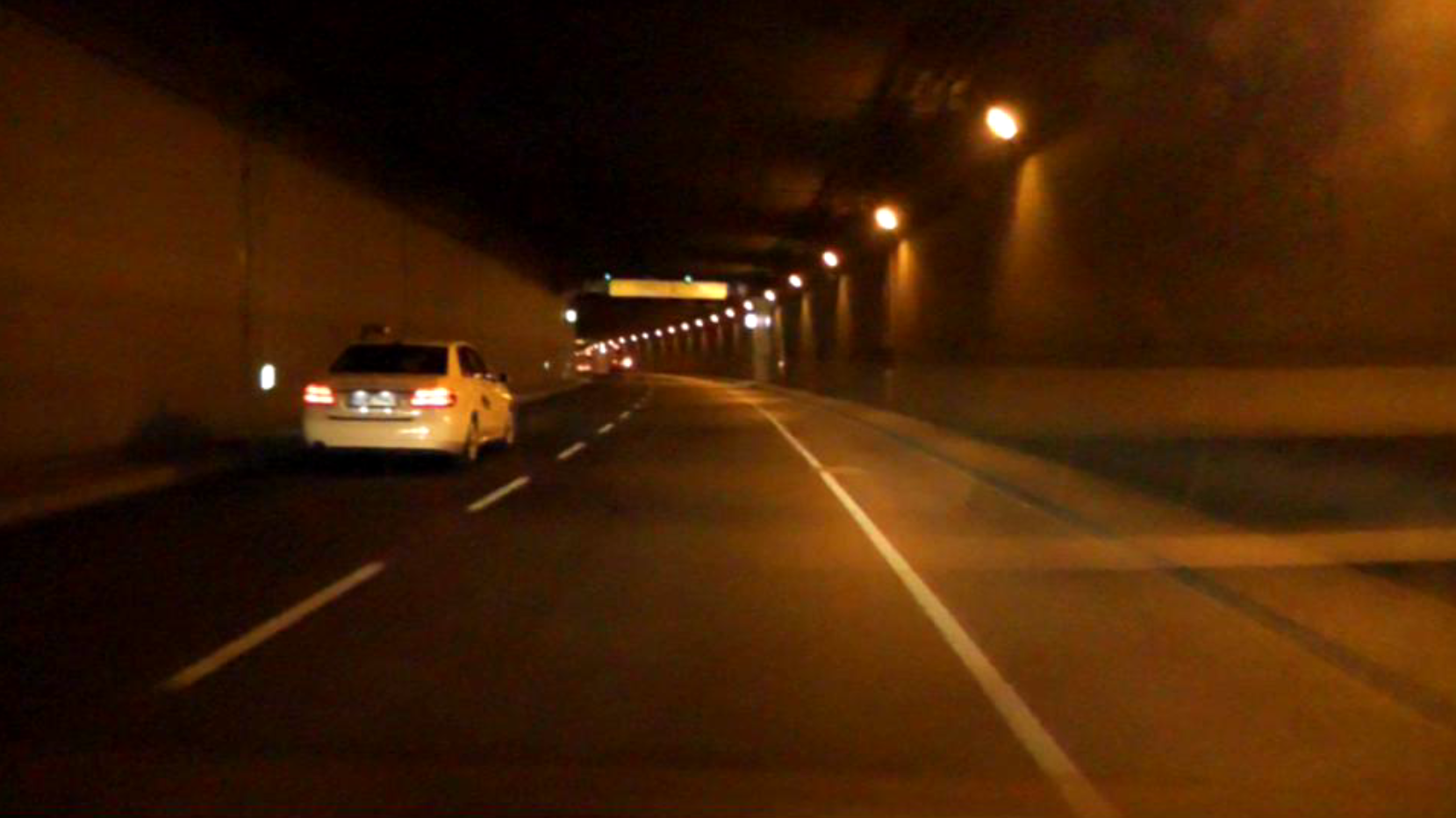}
\caption{Video snapshot of conducted measurement run in a tunnel scenario, taken from the rear measurement vehicle.}
\label{video}
\end{figure}

The tracking algorithms are applied to channel data from a measurement run in the so-called “Tiergartentunnel” in Berlin. The measurement vehicles are driving southbound in a convoy with speeds between 42 km/h and 52 km/h at distances between 75 m and 110 m, with the above mentioned uncertainty of around 20 m. The shape of the tunnel is curved and the traffic density during measurement was low, as can be seen in Fig. \ref{video}. One measurement run of the HHI channel sounder contains 10,000 snapshots, which are organized into sets of snapshots. In this measurement run, a set consists of six snapshots with a time interval of 0.717 ms between the snapshots, which results in a set recording time of around 3.6 ms. The time interval between the starting of two sets is 10 ms. We recorded 1666 sets, which amounts to a total measurement time of around 16.7 s. This measurement set-up enables longer measurement runs compared to continuous recording and in addition reflects the packet on-air times of cooperative awareness messages based on IEEE 802.11p. One of the most relevant challenges in vehicular communication is a reliable frame detection, which corresponds to the time-variant behavior of the channel while a transmission packet is on the air. Therefore, we selected the length of the set recording time to account for the maximum IEEE 802.11p frame duration of 2 ms, considering the maximum allowed payload of 1500 Bytes. 

\section{MPC Estimation and Tracking}
\subsection{Detecting MPCs in a Channel Impulse Response }
Processing of channel sounder data is a multi-step process and starts with the detection of MPCs in the channel impulse response at a certain time instance. Superresolution channel parameter estimation schemes such as ESPRIT \cite{roy89} and MUSIC \cite{schmidt86} require a covariance matrix, hence multiple observations of a stationary process. Since the fading process of vehicular channels is non-stationary \cite{bernado14}, \cite{he15} and \cite{bernado14} suggest to use a local WSS time window. Instead, we decided to use the smallest possible time window and take every measurement snapshot as an independent process step. This approach is reasonable for our wideband channel measurement data, since we observe non-stationary MPC behavior between snapshots, e.g. appearing/disappearing MPCs and a change of the MPC delay. Since our tracking method does not depend on any statistics and we are not aiming at any statistical channel parameters based on a local stationarity region, we disregard matters on appropriate stationarity window lengths in this paper.

We use a MPC detection algorithm that is based on work from \cite{santos08}, which has some similarities to the CLEAN \cite{Cramer02} or UWB-SAGE \cite{Haneda03} algorithms and can be summarized as follows: Find the strongest peak in the channel impulse response, subtract this dominant peak in the frequency domain and continue searching and subtracting strongest peaks respectively in the remaining impulse response. A detailed description of this search and subtract algorithm can be found in \cite{falsi06}, an experimental verification of this algorithm is given in \cite{santos08}. Here, we start with the time-variant channel impulse response as
\begin{equation}\label{gl:barwq}
h(t, \tau) = \sum^L_{l=1} \alpha_l(t) \delta ( \tau - \tau_l(t)),
\end{equation}
where $L$ is the number of scatterers, $\alpha_l$ the complex gain and $\tau_l$ the delay of MPC $l$. The channel impulse response provided from a channel sounder with measurement bandwidth $B$ can be expressed as
\begin{equation}\label{gl:barwq}
h(t, \tau) =  \sum^L_{l=1} \alpha_l(t) w( \tau - \tau_l(t)),
\end{equation}
where $w(\tau)$ is the isolated pulse with duration $T_p$. The sampled channel data depends on the snapshot sampling period $T_s$ and the delay resolution period $T_b = 1/B$. We therefore express the discretized form of the channel impulse response as 
\begin{equation}\label{gl:barwq}
h(n T_s, u T_b) =  \sum^L_{l=1} \alpha_l(n T_s) \: w( u T_b - \tau_l(n T_s)), 
\end{equation}
for $n=1,...,M$ and $u=1,...,U$, where $M$ is the maximum number of available snapshots and $U$ is the selected sounding sequence length (denoted $N$ in \cite{paschalidis08}). As previously stated, we want to detect the MPCs at each single snapshot $n$ and therefore define a channel impulse response as vector $\boldsymbol{h} \in \mathbb{R}^U$ with elements
\begin{equation}\label{gl:barwq}
h_u = h(u T_b) =  \sum^L_{l=1} \alpha_l \: w( u T_b - \tau_l).
\end{equation}
We furthermore define vector 
\begin{equation}\label{gl:barwq}
\boldsymbol{w}(\tau) = [ \boldsymbol{0}_{D_\tau} ,  \  \boldsymbol{w}_0 , \  \boldsymbol{0}_{U-Z-D_\tau}]^T  \in \mathbb{R}^U, 
\end{equation}
with $D_\tau$ being the discretized version of time delay $\tau$, such that $\tau \simeq D_\tau \cdot T_b$ and with $\boldsymbol{w}_0 \in \mathbb{R}^Z$ of elements $w_u = w(u T_b)$, $u=1,...,Z$ such that $T_p = Z \cdot T_b$. The zero series $\boldsymbol{0}_{D_\tau}$ and $\boldsymbol{0}_{U-Z-D_\tau}$ consist of ${D_\tau}$ and ${U-Z-D_\tau}$ zero elements respectively and are used to shift $\boldsymbol{w}_0$ in $\boldsymbol{w}(\tau)$. Since we are using sparse wideband channel data, we assume that the delay components $\tau_l$ are separable, that is,
\begin{equation}\label{gl:sep}
\left| \tau_i - \tau_j \right|  \geq T_b  \quad    \forall i \neq j
\end{equation}
and can therefore apply the maximum likelihood method to obtain the delay and complex amplitude estimates of the strongest peak \cite{falsi06}
\begin{equation}\label{gl:barwq}
\hat \tau_{l} = \underset{\tau}{\operatorname{argmax}} \left| \boldsymbol{w}(\tau)^T \boldsymbol{h}_{l}  \right|
\end{equation}
\begin{equation}\label{gl:barwq}
\hat \alpha_{l} = \frac{ \boldsymbol{w} (\hat \tau_{l})^T \boldsymbol{h}_{l} }{\boldsymbol{w}^T \boldsymbol{w}},
\end{equation}
where $\boldsymbol{h}_l$ is the channel impulse response after $l$ strongest peak detections. The strongest MPC is subtracted from the channel impulse response in the following way 
\begin{equation}\label{gl:minus}
\boldsymbol{h}_l =
  \begin{cases}
 \boldsymbol{h}      & \quad, l=1\\   
\boldsymbol{h}_{l-1}  - \hat \alpha_{l-1} \boldsymbol{w}^* (\hat \tau_{l-1})  & \quad ,l >1,\\
  \end{cases}
\end{equation}
the detected MPC $\theta_l = \{a_l, \tau_l \}$ is saved for further processing  and the algorithm searches for the next strongest peak.

We augment the algorithm by windowing the measured transfer function before applying the detection algorithm. The windowing reduces the side lobes of the pulses in the channel impulse response and improves the overall detection performance. In order to select the best windowing method, the performance evaluation described in Section IV.A was executed for a Raised-Cosine roll-off window and a Kaiser window \cite{kaiser80} with different corresponding window parameters. The shape of the Kaiser window in the frequency domain
\begin{equation}\label{gl:28}
z[u]=\begin{cases}\frac{ I_0 \left( \pi a \sqrt{ 1 - (\frac{2u}{U-1} -1)^2} \right )  }{ I_0 ( \pi a )}, & \text{$0 \le u \le U-1 $}\\
0 &\text{otherwise,} \end{cases}
\end{equation}
and therefore the trade-off between the width and the side lobes of the pulse is determined only by   one parameter $a$. In (\ref{gl:28}), $I_0$ is the zeroth order of the modified Bessel function of the first kind, $U$ is the window length (equal to the sounding sequence length) and $a$ a non-negative real number. The best results showed the Kaiser window with a parameter value of $a$ = 6.

After the subtraction of a peak, a delay range around the identified peak location is blocked for the following peak searches, with the purpose to prevent a re-detection at neighboring delay values. We set the width of this blocked delay range equal to the channel sounder pulse width at 10 dB below its peak magnitude, which is in our case 2.47 ns. The entire detection process for a channel impulse response involves the following steps: 
\begin{enumerate}
\item Estimate the noise floor by estimating the power of a channel impulse response part where no MPCs occur (usually at larger delays), add 6 dB to obtain a noise floor threshold and set all values in the impulse response below this threshold to zero.
\item Apply windowing in the frequency domain using a Kaiser window with a parameter value of 6 (see (\ref{gl:28}) for the corresponding equation and \cite{kaiser80} for more details).
\item Find the strongest peak outside blocked delay range(s) and save as a detected MPC.
\item Block the delay range surrounding the newly detected MPC.
\item Subtract the channel sounder pulse at the detected MPC delay position from the measured transfer function, as done in (\ref{gl:minus}).
\item Repeat points 3 to 5 until no additional MPCs are detected.
\end{enumerate}

\begin {table}
\caption {Parameters of the GM-PHD filter implementation used for tracking performance comparison} \label{tab:title} 
\begin{center}
 \begin{tabular}{|c c c|} 
 \hline
 Symbol & Value &  Explanation  \\ [0.5ex] 
 \hline\hline
 $\sigma_{\nu}$ & $3 \cdot 10^{-1}$ & Standard deviation of process noise in $\frac{m}{s^2}$ \\ 
 \hline
 $\sigma_{\varepsilon}$ & $3 \cdot 10^{-3}$ &  Standard deviation of measurement noise in $m$ \\
 \hline
 $p_{S}$ & 0.99 & Probability of target survival, see (19) in \cite{vo06}  \\
 \hline
 $p_{D}$ & 0.95 & Probability of target detection, see (20) in \cite{vo06}  \\
 \hline
 $T$ & $1 \cdot 10^{-2}$ & Truncation threshold, see Table II in \cite{vo06} \\
 \hline
 $U$ & $1 \cdot 10^{-1}$ & Merge threshold, see Table II in \cite{vo06} \\ 
 \hline
 $-$ & $2 \cdot 10^{-1}$ & Minimum weight threshold, see Table III in \cite{vo06} \\ [1ex] 
 \hline
\end{tabular}
\end{center}
\end {table}
\subsection{Tracking of MPCs over Time}
The large measurement bandwidth of 1 GHz makes MPC tracking of individual MPCs comparatively easy and leads to a reduced likelihood of false positives. In addition, a large bandwidth decreases the number of physical MPC in a superimposed MPC and consequently decreases the amplitude fluctuations. This can be confirmed with the short-term tracking results of our measurement run, where 90\% of the MPCs show a power standard deviation of less than 1.5 dB, as shown in the results section in Fig. \ref{StdPower}. Another fact that reduces the demands on an (ultra-) wideband MPC tracking algorithm in vehicular communications scenarios is the comparatively high predictability of the geometry changes; cars usually drive within clearly defined limits regarding their change of direction and change in speed. The high predictability of the moving objects results in a high predictability of the corresponding MPC tracks. Our wideband channel data reveals appearing and disappearing MPCs from one snapshot to the next snapshot as shown in Fig. \ref{MPCTracking}. Therefore, instead of defining a stationarity window across multiple measurement snapshots as done in \cite{he15}, we decided to use a time window equal to the time interval between two snapshots and track MPCs on a snapshot basis.

The goal of our MPC tracking method is to keep the algorithm complexity as low as possible and yet establish an effective tool with a good tracking performance. In order to evaluate our algorithm, we compare performance indicators with a state-of-the-art tracking algorithm called Gaussian mixture probability hypothesis density (GM-PHD) filter \cite{vo06}, which incorporates the widely used extended Kalman filter \cite{salmi09}. The GM-PHD filter is a recursive algorithm that models targets as random finite sets and propagates the posterior intensity in time. The implementation of this tracking approach is elaborate, requires substantial computational efforts and an adaption of least 10 algorithm parameters to the corresponding tracking problem. This includes seven core algorithm parameters and additional parameters with lower effect on the tracking results. In our implementation, we use three additional parameters for creating the so-called birth processes and four additional parameters to filter out tracks with unlikely delay/magnitude changes (similar to our maximum delay/magnitude change thresholds in (\ref{gl:maxchange})). In Table I we list the most relevant algorithm parameters matched to our measurement data.\\ 

Our proposed algorithm works with little computational effort and is based on the continuity of the delay and magnitude changes. Although proposed as a method for MPC tracking in \cite{meifang}, the phase change estimates are not used for tracking due to their high measurement noise and their $2\pi$-ambiguity. Our algorithm is based on four parameters, which are estimated depending on the measurement data set-up and quality as for (\ref{gl:erroreps}) or directly from the underlying physical model as for (\ref{gl:maxchange}). 

The developed MPC tracking algorithm is based on the following assumptions: 
\begin{enumerate}
\item The same MPC is detected in three consecutive snapshots.
\item Splitting or combining of MPC tracks does not occur. 
\item The second derivatives of delay and magnitude, i.e. the change of the Doppler frequency (delay change) and the change of the MPC power fading (magnitude change), are below the estimated maximum search tolerances in (\ref{gl:erroreps}).
\end{enumerate}

\begin{figure}[!t]
\centering
\includegraphics[width=3.3in]{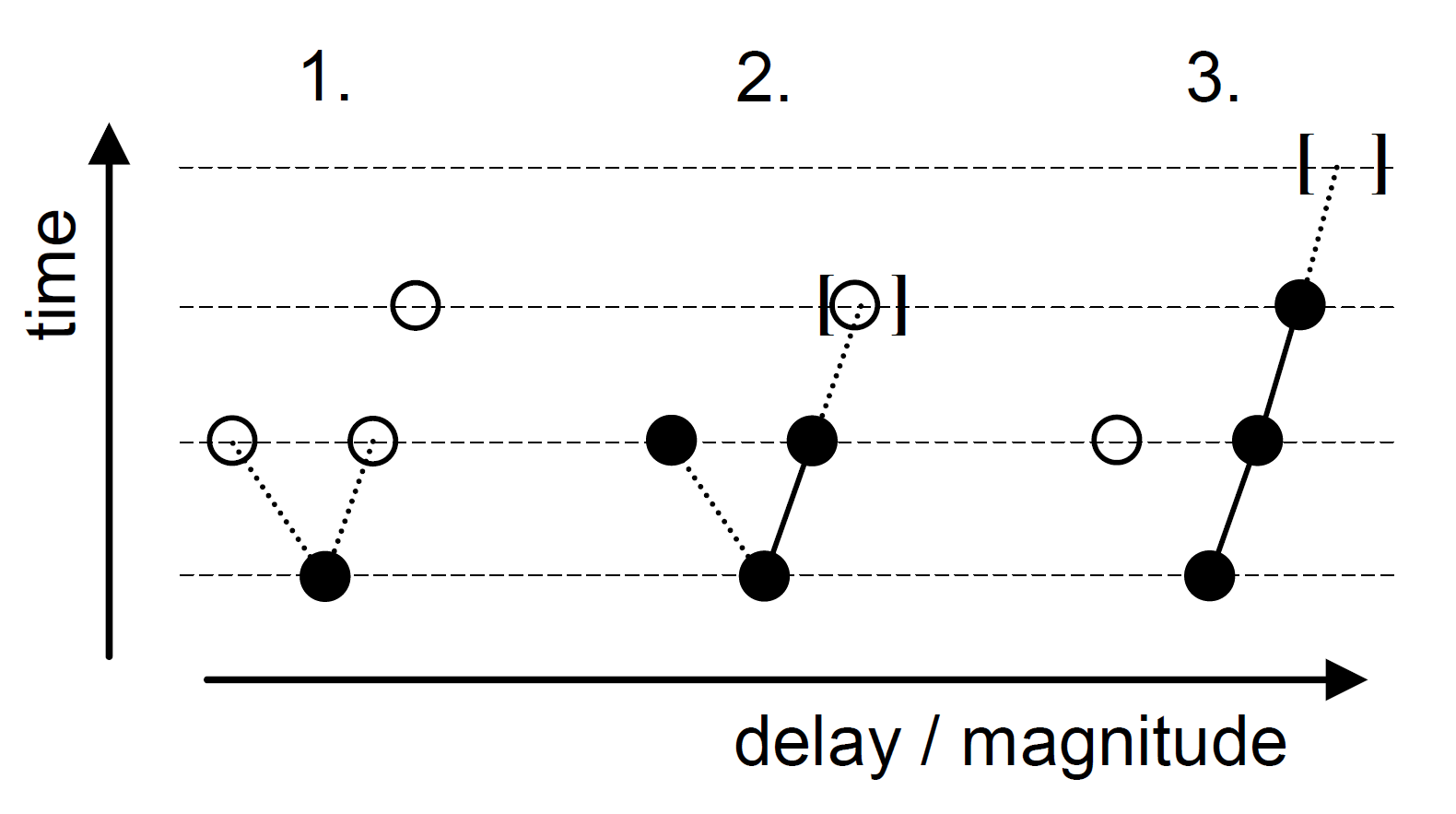}
\caption{Schematic diagram of the MPC tracking steps.}
\label{algo}
\end{figure}

The main idea behind our MPC tracking approach is similar to the tracking approach described in \cite{karedal09}, but in addition to delay estimates we also use magnitude estimates as a measure for tracking. The MPC tracking steps are depicted in Fig. \ref{algo} and can be summarized as follows:
\begin{enumerate}
\item Start in the first snapshot with the strongest peak and search in the second snapshot for neighboring peaks in terms of delay and magnitude distance to the starting peak. 
\item Use the observed delay change and magnitude change (dotted lines in step 1 in Fig. \ref{algo}) to predict the peak location in the third snapshot, as shown in step 2 in Fig. \ref{algo}. Based on the predicted location and the pre-defined search tolerances, define a two-dimensional search range (brackets in step 2 in Fig. \ref{algo}). If a peak is found within the search ranges, a MPC track has been identified.
\item In case more than one peak is found within the defined search ranges, choose the peak with smallest distance to the predicted delay location.
\item Use the latest delay change and magnitude change to get the next search ranges accordingly (step 3 in Fig. \ref{algo}). Continue searching peaks along the MPC track, until no peak within the current search ranges is found.
\item Start with the second strongest peak in the first snapshot until the MPC track ends. Only consider peaks that are not yet linked to a MPC track.
\item Continue searching tracks for all other peaks in the first snapshot and then continue searching for tracks in later snapshots that are not yet linked to a MPC track.
\end{enumerate}

Stating the MPC tracking algorithm with mathematical expressions, we have to keep in mind that $h(nT_s,uT_b) = h[n,u]$, with $T_s$ being the channel acquisition period and $T_b$ being the delay resolution period. The output from the detection algorithm at time instance $n$ are the MPC gain $a_l$, the delay $\tau_l$ and the phase $\phi_l$
\begin{equation}\label{gl:barwq}
\theta_l = \{a_l, \tau_l, \phi_l \}
\end{equation}
with $l  \in \{1...L_n\}$ and $L_n$ being the number of detected MPCs at time $n$. For the MPC tracking algorithm, we use a subset of the estimated parameters
\begin{equation}\label{gl:barwq}
\boldsymbol{s}_l[n] = \left( \begin{array}{c} a_l[n] \\ \tau_l[n] \end{array} \right) .
\end{equation}
We start in $n=n_k^{start}=1$ searching for MPC track $k=1$ with the strongest MPC $\boldsymbol{\hat s}[n]$ and its neighbors in the next time instance
\begin{equation}\label{gl:8}
x_l^c=\begin{cases}1, & \text{if $d \left(\boldsymbol{\hat s}[n],\{ \boldsymbol{s}_l[n+1]\}_{l=1}^{L_{n+1}} \right) \leq \boldsymbol{\xi}_s$}\\
0 &\text{otherwise,} \end{cases}
\end{equation}
where $d(.,.)$ is the distance defined as
\begin{equation}\label{gl:9}
d \left( \boldsymbol{\hat s}[n],\{ \boldsymbol{s}_l[n+1]\}_{l=1}^{L_{n+1}} \right) = \left| \left( \begin{array}{c} \hat a[n]-\{a_l[n+1]\}_{l=1}^{L_{n+1}} \\  \hat \tau_l[n]-\{\tau_l[n+1]\}_{l=1}^{L_{n+1}} \end{array} \right) \right|,
\end{equation}
and 
\begin{equation}\label{gl:maxchange}
\boldsymbol{\xi}_s = \left( \begin{array}{c}\xi_a \\ \xi_\tau
\end{array} \right)
\end{equation}
is the maximum magnitude change and delay change, based on considerations on the physical limits of the moving objects. The resulting $C$ initial track direction candidates 
\begin{equation}\label{gl:barwq}
 x_l^c \{ \boldsymbol{s}_c[n+1]\}_{c=1}^{C} = \{\boldsymbol{s}^1_l[n+1],\boldsymbol{s}^2_l[n+1],...,\boldsymbol{s}^{C}_l[n+1]\}
\end{equation}
are used for an identification of track $k$ by predicting $\tilde s_k[n+1]$ with the linear prediction model $H$. We start with the first initial direction candidate and set $s_l[m] = s_l^1[n+1]$ in
\begin{equation}\label{gl:12}
\boldsymbol{ \tilde s}_k[m+1] = H\left( \boldsymbol{s}_l[m] \right)  = \left( \begin{array}{c} a_l[m]+\Delta a_l [m] \\ \tau_l[m]+\Delta \tau_l [m]
\end{array} \right),
\end{equation}
with
\begin{equation}\label{gl:barwq}
\Delta a_l[m] = a_l[m] - a_l [m-1]
\end{equation}
\begin{equation}\label{gl:barwq}
\Delta \tau_l[m] = \tau_l[m] - \tau_l [m-1].
\end{equation}
Based on the prediction $\boldsymbol{\tilde s}_k^1[n+2] = \boldsymbol{\tilde s}_k[m+1]$, we can look for MPCs in the defined search ranges $\boldsymbol{r}_{min}$ and $\boldsymbol{r}_{max}$ 
\begin{equation}\label{gl:barwq}
x_{kl}^{m}=\begin{cases}1, & \text{if $\boldsymbol{r}_{min}  \leq\{ \boldsymbol{s}_l[m+1]\}_{l=1}^{L_{m+1}} \leq \boldsymbol{r}_{max}$}\\
0 &\text{otherwise,} \end{cases}
\end{equation}
 where
\begin{equation}\label{gl:barwq}
\boldsymbol{r}_{min} = \boldsymbol{\tilde s}_k[m+1]-\boldsymbol{\epsilon}_s
\end{equation}
\begin{equation}\label{gl:barwq}
\boldsymbol{r}_{max} = \boldsymbol{\tilde s}_k[m+1]+\boldsymbol{\epsilon}_s.
\end{equation} 
The values of the maximum allowed search tolerances 
\begin{equation}\label{gl:erroreps}
\boldsymbol{\epsilon}_s = \left( \begin{array}{c}\epsilon_a \\ \epsilon_\tau
\end{array} \right)
\end{equation}
depend on the time interval between two snapshots, the dynamics of the propagation channel and the quality of the measurement device. An evaluation of the algorithm results has to be performed in order to estimate appropriate search tolerance values, as described in Section IV.B.

Then, we check the outcome of the binary variable $x_{kl}^{m}$ with
\begin{equation}\label{gl:20}
\sum^{L_{m}}_{l=1} x_{kl}^{m}= X. 
\end{equation}
In case $X=0$, no track is found and the algorithm continues with the next \textit{initial candidates} $ \{ s^c_l[n+1]\}$. In case  $X=1$, only one \textit{track candidate} is found and consequently a track is identified; while in case $X>1$, more than one track candidate is found. In order to select from multiple track candidates, we use
\begin{equation}\label{gl:barwq}
\underset{x_{kl}}{\operatorname{argmin}}\sum^{L_{m+1}}_{l=1}d(\tilde \tau_k[m+1],\tau_l[m+1]) x_{kl}
\end{equation}
so that $X=1$ holds in  (\ref{gl:20}). Now, as an MPC track is identified by three adjacent MPCs, we save  
\begin{equation}\label{gl:barwq}
 x_{kl}^{n} = x_{kl}^{n+1}=x_{kl}^{n+2}= 1
\end{equation}
and continue searching along track $k=1$ with the linear prediction model  in (\ref{gl:12}), for  $m=n_k^{start}+3 , ...,  N$ or until the end of track $k$ at $N_k$ is found. We save the estimated MPC parameters of track $k$ in
\begin{equation}\label{gl:barwq}
\boldsymbol{s}_k[n] = \boldsymbol{s}_l[n] x^n _{kl}.
\end{equation}
Next, we continue in $n=n_k^{start}=1$ with the next strongest MPC $\hat s[n]$ not yet part of a track
\begin{equation}\label{gl:25}
\boldsymbol{\hat s}[n] \notin \{ \boldsymbol{s}_k[n]\}_{k=1}^{K_{n}},
\end{equation}
where $K_n$ is the number of tracked MPCs at time $n$. Based on $\boldsymbol{\hat s}[n]$, we start again identifying initial track direction candidates in (\ref{gl:8}). After $L_n$ starting MPCs $\boldsymbol{\hat s}[n]$ are considered, we continue with $n=n_k^{start} = 2$ and look for MPCs that fulfill (\ref{gl:25}) to start again from (\ref{gl:8}). The result of the tracking algorithm is the set
\begin{equation}\label{gl:barwq}
\boldsymbol{S}_k = \{ \boldsymbol{s}_k[n]\}_{n=n_k^{start}}^{N_{k}}.
\end{equation}
The lifetime of track $k$ is $\psi_k = N_k - n_k^{start}$. In order to obtain a Doppler frequency estimate per track $k$, we first retrieve $m_k$ with linear regression
\begin{equation}\label{gl:barwq}
y + m_k x = \frac{\Delta\{\tau_k[n]\}_{n=n_k^{start}}^{N_{k}} }{\Delta n T_s}
\end{equation}
and calculate the Doppler frequency
\begin{equation}\label{gl:29}
\nu_k = -m_k f_c
\end{equation}
where $f_c$ is the carrier frequency. The final outcome of the short-term MPC tracking are the following estimates
\begin{equation}\label{gl:barwq}
\theta_k = \{a_k[n], \tau_k[n], \psi_k, \nu_k \}_{n=n_k^{start}}^{N_{k}}.
\end{equation}

\section{MPC Tracking Performance Evaluation }

\subsection{Evaluation Based on an Artificial Channel }
\begin{figure}[!t]
\centering
\includegraphics[width=3.3in]{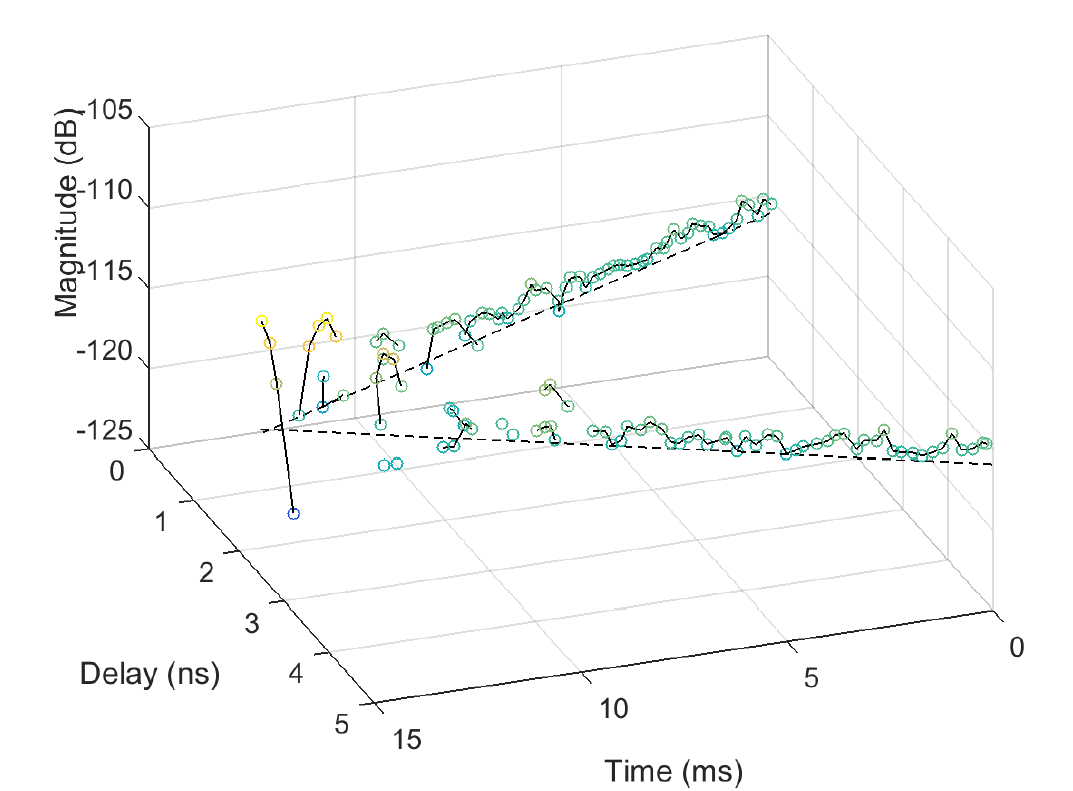}
\caption{Result from proposed MPC tracking algorithm for an artificial channel created with a measured channel sounder pulse.}
\label{ApproachingTracks1}
\end{figure}

In order to evaluate the performance of the detection algorithm and both tracking algorithms, two different kinds of evaluation methods are applied. The first evaluation method is based on an artificial channel, which is created using a channel sounder pulse. This pulse is extracted from a channel sounder measurement via an RF cable and reflects the characteristics of the measurement device. The artificial channel used for the performance evaluation consists of two MPCs tracks with identical power levels, decreasing delay distances and an added noise floor with a power level at -140 dB, as shown in Fig. \ref{ApproachingTracks1}. This artificial channel can be regarded as a worst-case tracking scenario when considering two MPCs, since a greater power difference between the MPCs would lead to the dominance of one MPC and a better overall result. Since we are mainly interested in the separation of two MPC tracks that are close in terms of delay distance, we use this simple artificial channel to assess the tracking performance.

\begin{figure}[!t]
\centering
\includegraphics[width=3.4in]{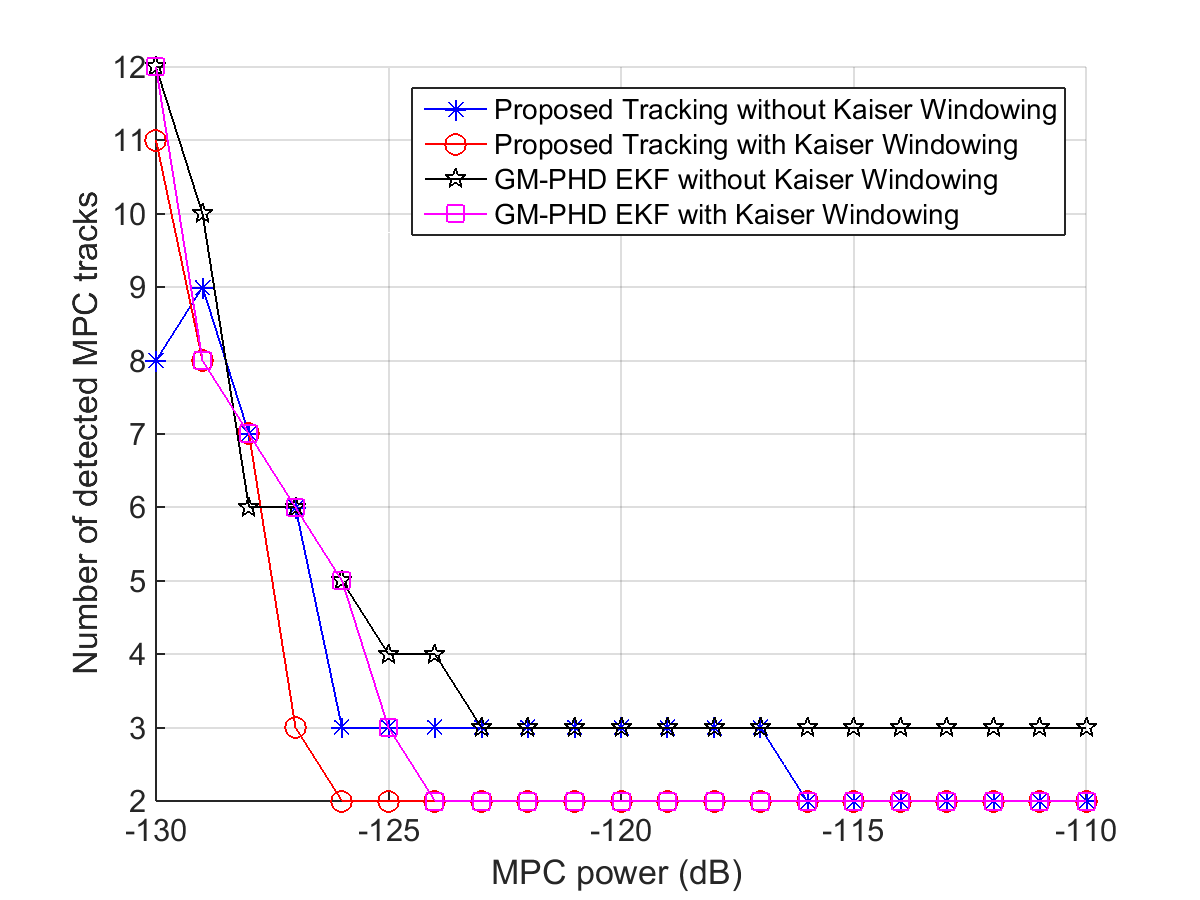}
\caption{Number of detected tracks as a MPC tracking performance indicator, based on an artificial channel and delay distances above 2.5 ns}
\label{NumDetectTracks}
\end{figure}

\begin{figure}[!t]
\centering
\includegraphics[width=3.4in]{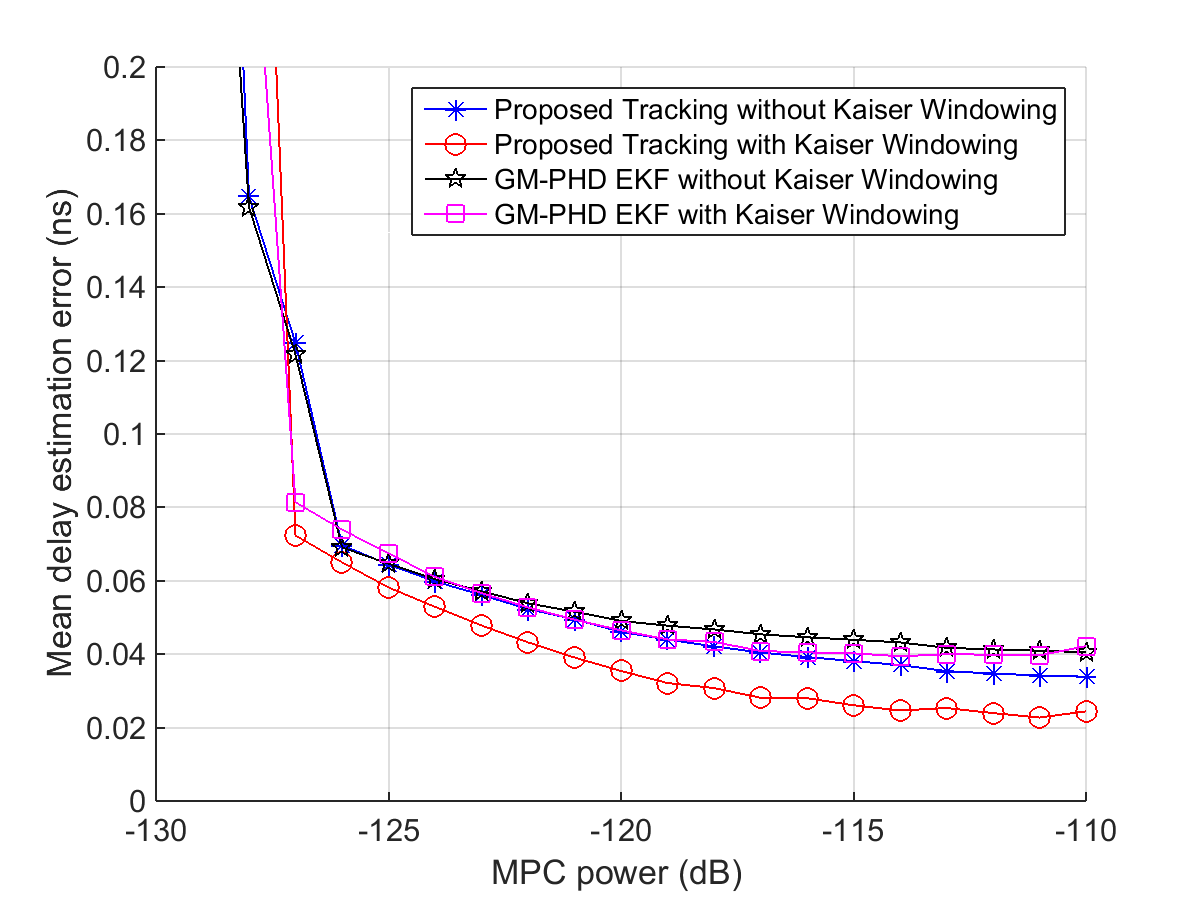}
\caption{Mean delay estimation error as a MPC tracking performance indicator, based on an artificial channel and delay distances above 2.5 ns}
\label{DelayEstimateError}
\end{figure}

The example in Fig. \ref{ApproachingTracks1} shows an artificial channel with MPC powers of -116 dB. We found that up to a delay distance of around 2.5 ns, the mean delay estimation error is below 0.03 ns with flawless tracking results, as shown for large MPC powers in Fig. \ref{NumDetectTracks} and Fig. \ref{DelayEstimateError}. The malfunction of the detection algorithm at delay distances below 2.5 ns can be explained with the width of the channel sounder pulse and the effects of two superimposing pulses \cite{Fleury99}. The decreasing distance of two complex pulses with finite bandwidth leads to constructive or destructive superposition and consequently to fluctuations of the resulting pulse, as shown in Fig. \ref{ApproachingTracks1}. In order to find the limits of the applied algorithms, the power level of the MPCs is reduced and the tracking results up to a distance of 2.5 ns between the approaching MPC tracks are compared to ground truth. 

We compare the tracking performance of the proposed algorithm and the GM-PHD filter in Fig. \ref{NumDetectTracks} and Fig. \ref{DelayEstimateError}, both with and without prior windowing. As we can observe in Fig. \ref{NumDetectTracks} and Fig. \ref{DelayEstimateError}, prior Kaiser windowing results in more accurate number of tracks and lower mean delay estimation errors. The third appearing "track" without Kaiser windowing in Fig. \ref{NumDetectTracks} is due to the fact that the superimposing pulses generate a third pulse in some successive snapshots, which are misinterpreted as a track. Also, we find in Fig. \ref{NumDetectTracks} that the detected MPC tracks start to split into more than the actual two tracks at a MPC power of -124 dB for the GM-PHD filter and at a MPC power of -126 dB for the proposed algorithm. As can be found in Fig. \ref{DelayEstimateError}, the mean delay estimation error at these power values is below 0.07 ns and in terms of delay estimation, the proposed tracking algorithm has a better performance than the GM-PHD filter.

\begin{figure}[!t]
\centering
\includegraphics[width=3.4in]{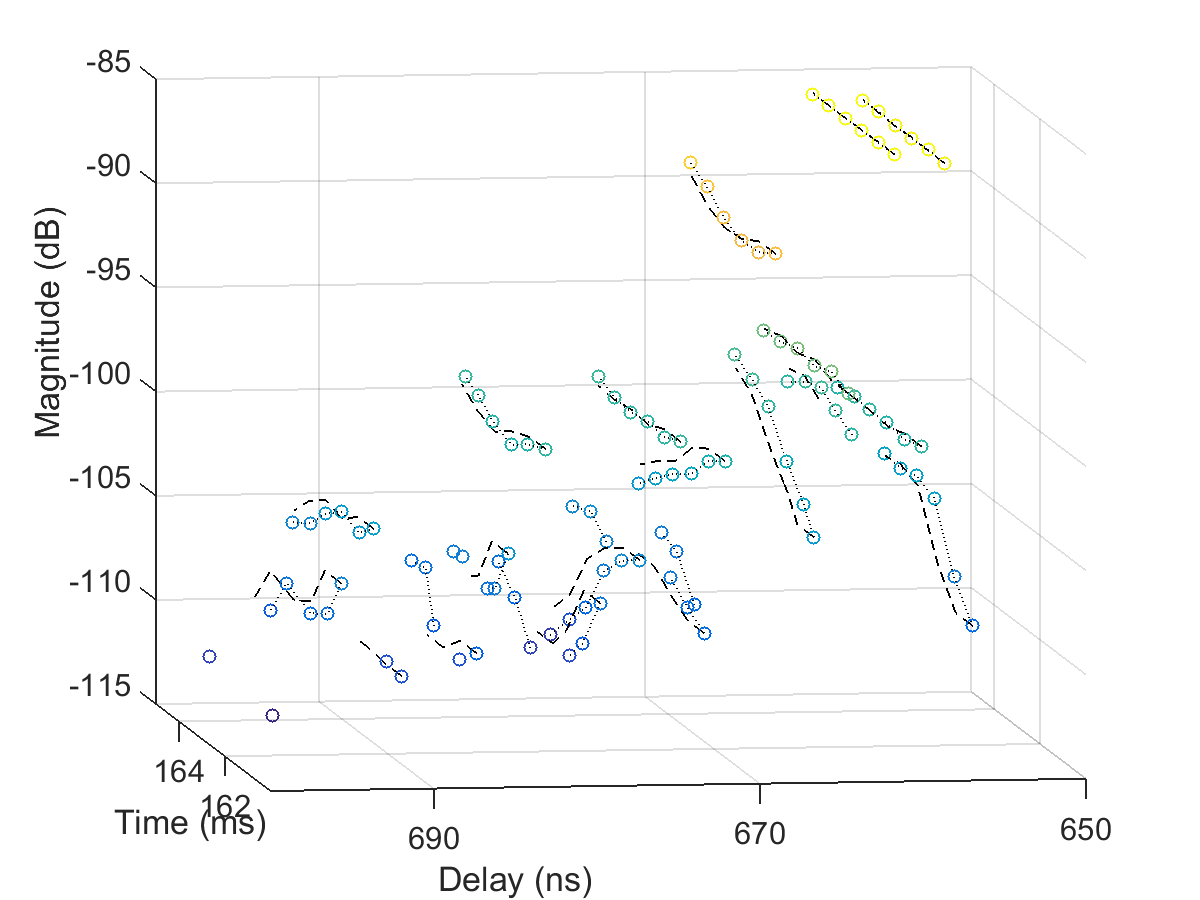}
\caption{MPC tracking results for a recorded channel measurement set in a tunnel convoy traffic communication scenario. The dashed lines result from the GM-PHD filter with EKF, whereas the dotted lines result from the proposed tracking algorithm}
\label{MPCTracking}
\end{figure}

\begin{figure}[!t]
\centering
\includegraphics[width=3.4in]{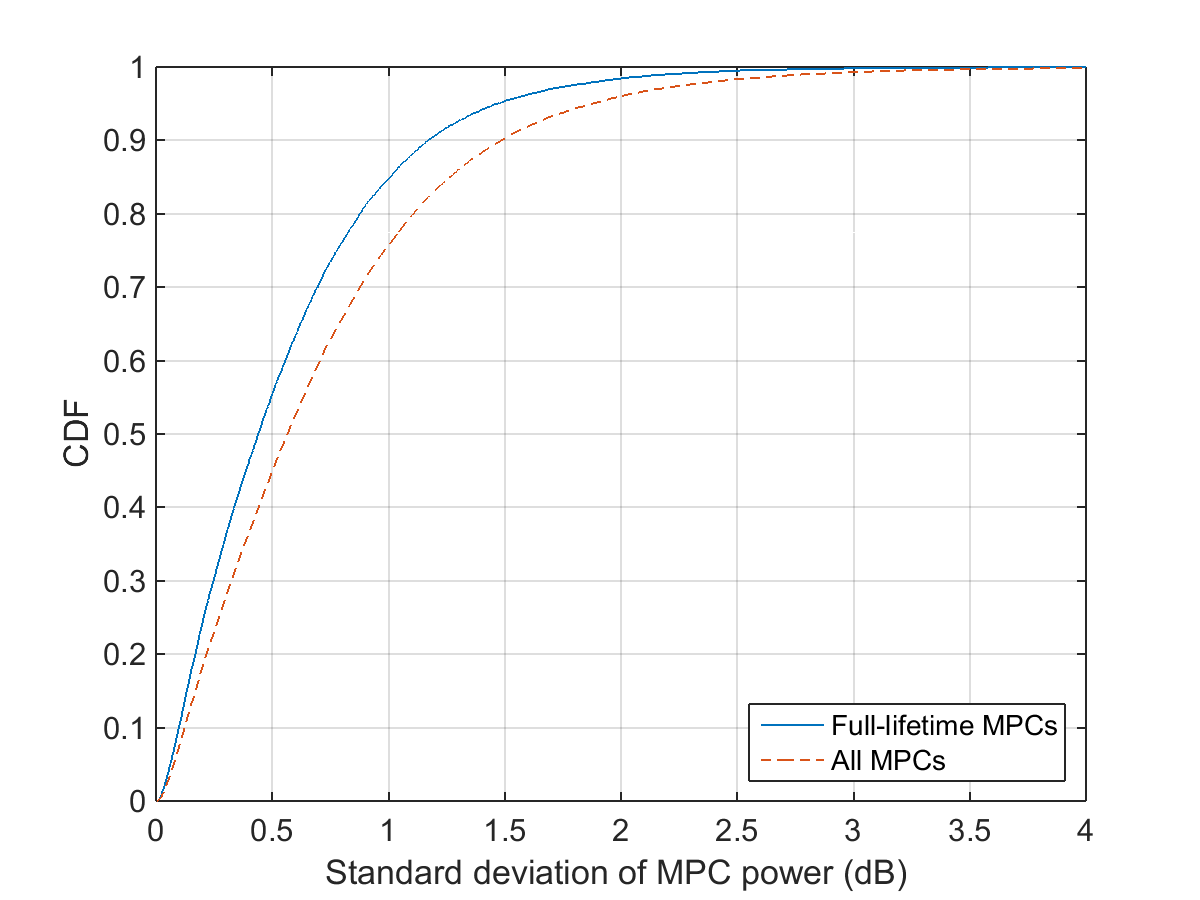}
\caption{Standard deviation of tracked wideband multipath components power for entire measurement run}
\label{StdPower}
\end{figure}

\begin{figure}[!t]
\centering
\includegraphics[width=3.4in]{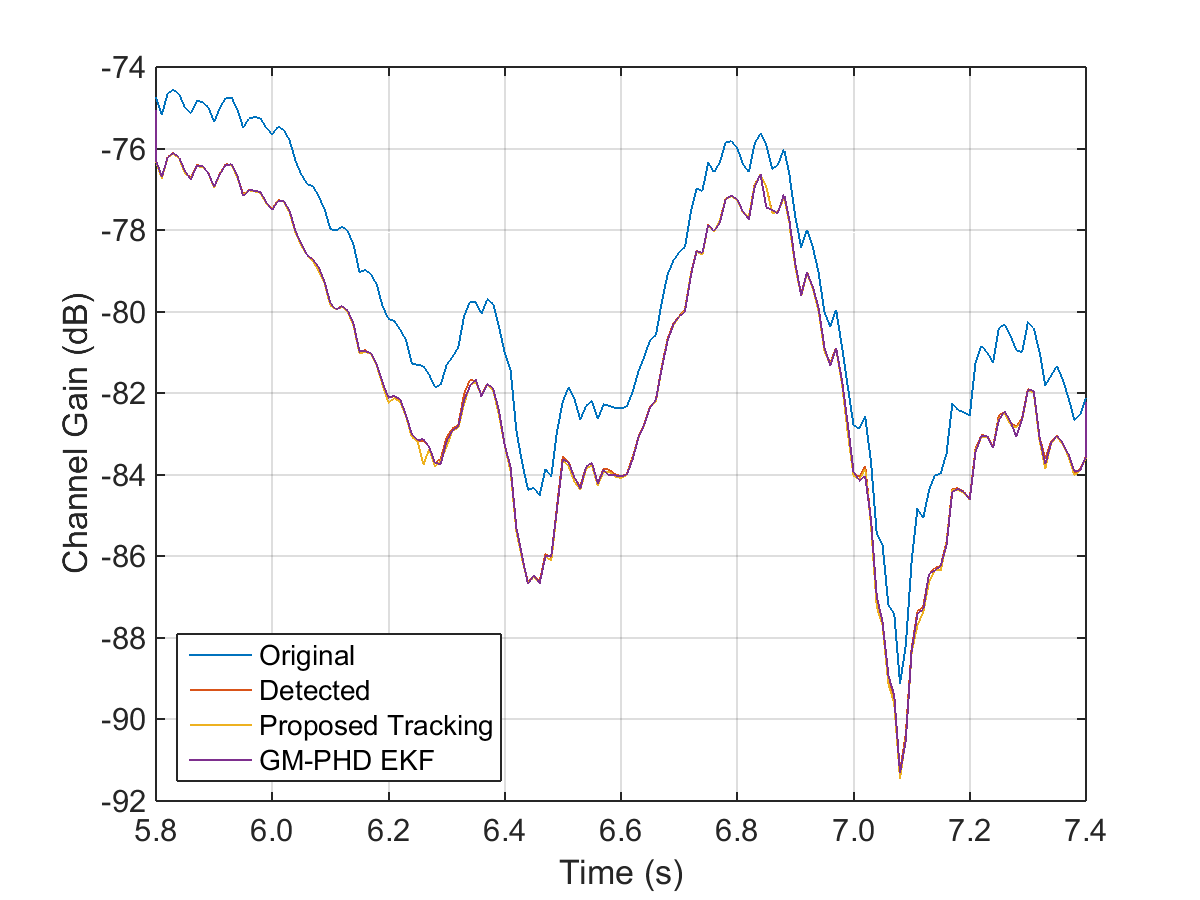}
\caption{Comparision between the original channel gain after Kaiser windowing and the captured channel gain after the detection, the proposed tracking algorithm and the GM-PHD filter with EKF}
\label{TrackedEnergy}
\end{figure}

\begin{figure}[!t]
\centering
\includegraphics[width=3.4in]{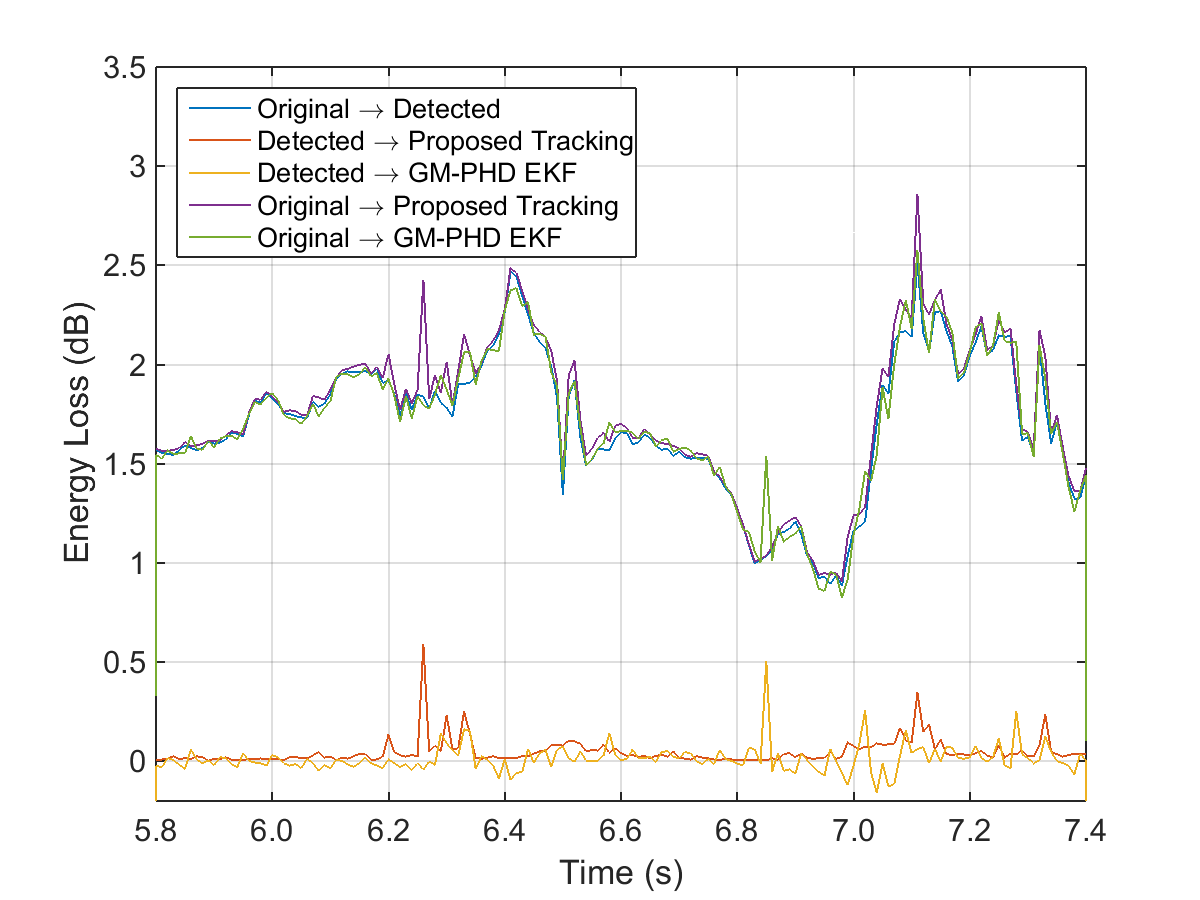}
\caption{Power losses due to the detection, the proposed tracking algorithm and the GM-PHD filter with EKF}
\label{DiffTrackedEnergy}
\end{figure}

\subsection{Evaluation Based on Channel Measurement Data}
The second performance evaluation is based on visual inspection of tracking results from actual channel measurement data. The goal of this evaluation is to identify false positives and false negatives, i.e. tracking mistakes and missed tracks. In order to evaluate the tracking algorithms on demanding MPC tracks, we use the previously described measurement run from a convoy traffic tunnel scenario with dense multipath inter-arrival times. Based on the results, we tuned the parameters of the tracking algorithm empirically and found that with a snapshot time interval of  $T_s$ = 0.7 ms, a delay search tolerance value of $\epsilon_\tau$ = 0.5 ns and a magnitude search tolerance value of $\epsilon_a$ = 10 dB is suitable. These high search tolerance values have been selected since they leave enough space for deviations from the instantaneously observed delay/magnitude change without, due to the sparsity of the channel, leading to additional false positives. In particular weak MPCs undergo significant fluctuations and require larger search tolerances. These search tolerance values are most likely also adequate for other V2V scenarios, if applied on measurement data with similar snapshot time intervals and similar measurement bandwidth. This assumption is based on the fact that these values were also appropriate for a highway measurement run we processed and analyzed.

Fig. \ref{MPCTracking} shows the time-variant channel impulse response of a tunnel scenario measurement, where circles indicate the detected MPC peaks, the dotted lines show the outcome of the proposed tracking algorithm and the dashed lines show the results from the GM-PHD filter. In Fig. \ref{MPCTracking}, we can observe that, compared to the GM-PHD filter tracks, the tracks from the proposed algorithm are more reactive to the dynamic MPC behavior. Also, we can observe that some diffuse multipath components (DMC) peaks are not linked to any MPC track and result in a loss of the captured channel gain. Fig. \ref{TrackedEnergy} shows for a part of our measurement run the original channel gain after the mentioned Kaiser windowing, the detected channel gain and the tracked channel gain of both tracking approaches, which are defined at time instance $n$ by $\sum_{j=1}^J |a_j|^2$, $\sum_{l=1}^{L_n} |a_l|^2$ and $\sum_{k=1}^{K_n} |a_k|^2$ respectively. It can be observed that all processed channel gain curves, the detected and the two tracked curves, are below the original channel gain curve and follow the original curve progression very well with a certain distance. 

Fig. \ref{DiffTrackedEnergy} displays the power losses due to the applied algorithms, where it becomes clear that the main loss is due to the detection algorithm. Also, we can observe in Fig. \ref{DiffTrackedEnergy} that the loss decreases for a larger channel gain, which can be explained with DMCs being beyond the limited dynamic range of the channel sounder when receiving stronger MPCs. When comparing the two tracking approaches in Fig. \ref{DiffTrackedEnergy}, we can observe that each algorithm has tracking difficulties at different parts of the measurement run (the proposed algorithm at 6.25 s and the GM-PHD filter at 6.85 s).
Furthermore, we can observe that the energy loss of the GM-PHD filter is sometimes below zero. This can be explained with the set-up of this algorithm and the "survival" of a track, even without any measurement data supporting this track. In contrast, the proposed algorithm simply connects detected MPCs and therefore never results in a negative loss. The mean square error of the energy loss compared to the detected channel gain is 0.0065 dB for the proposed algorithm and 0.005 dB for the GM-PHD filter. For the entire measurement run, we observed an average total energy loss of around 2 dB for both tracking approaches, with a standard deviation of around 0.4 dB. 

The main advantage of our proposed algorithm is its comparatively low computational effort. The performance of our tracking method is comparable to the GM-PHD filter, but its numerical complexity is strongly reduced. The proposed algorithm has a linear time complexity, due to the fact that it processes on a snapshot basis every MPC track separately. In contrast, the GM-PHD filter predicts and processes multiple targets simultaneously, which leads to a linear complexity in the number of targets and a cubic complexity in the number of snapshots \cite{Ba07}. In order to limit the computational effort of a corresponding implementation, several thresholds are used to reduce the number of targets (e.g. the last three variables in Table I). However, strongly ``optimized'' threshold values come at the expense of a reduced tracking accuracy. 

We estimate a time complexity reduction factor of up to 10, based on the fact that processing 100 measurement snapshots with our algorithm take around 95 s on a standard computer, whereas processing with our GM-PHD filter implementation requires 911 s.
    
In addition, our algorithm leads to  more accurate delay estimates if applied on measurement data with low measurement noise, as can be found in Fig. \ref{DelayEstimateError}. This is due to the fact that our approach interrelates detected peaks instead of generating processes that approximate tracks, as done in the GM-PHD filter. In other words, the GM-PHD filter usually lags a little behind the measured dynamics. Furthermore, since the resulting track is not directly linked to the detected peaks, a track might be found where there is actually no track. Finally, due to the high number of algorithm parameters, finding the proper parameter value set is a challenging task. We do not claim that our algorithm is applicable for any tracking problem, but it showed that this approach is a better solution for our measurement data compared to the GM-PHD filter. One disadvantage of our proposed algorithm is that it fails to track if multiple peaks in adjacent snapshots are missing. This shortage could be overcome by continuously searching for peaks along the observed delay (and magnitude) change, as done in \cite{karedal09}. On the other hand, bridging a track across multiple snapshots might result in less reliable tracking results and furthermore does not reflect what has been measured. 

From Fig. \ref{MPCTracking}, we can observe that the power of strong MPCs stay nearly constant within a measurement set, whereas smaller MPC tracks show larger power fluctuations and shorter lifetimes. An analysis of the MPC power standard deviation over the entire measurement run results in a CDF as shown in Fig. \ref{StdPower}, based on the proposed tracking results of 34,000 MPC tracks and 23,000 full-lifetime MPC tracks respectively. The long-term MPC tracking algorithm described in the next section only takes full-lifetime MPCs into consideration, where 85\% of the MPCs show a power standard deviation of less than 1 dB over the set recording period of around 3.6 ms.

\section{Long-term Tracking }
In order to investigate the large-scale evolution of MPC tracks, a supplementary long-term tracking algorithm is applied. This additional tracking method is needed to interrelate MPC tracks across adjacent recording sets, which are separated by gaps of around 6.4 ms as described in Section III.B. For this algorithm, only full-lifetime MPCs are considered, i.e. MPC tracks with a lifetime equal to the duration of the recording set. Excluding non-full-lifetime MPC tracks is based on the observation that tracks appearing or disappearing within the set recording time of 3.6 ms rarely lead to a MPC survival of 10 ms or more. Disregarding these MPC tracks results in an additional power loss of around 5\%, but increases the reliability of the long-term tracking.

\begin{figure}[!t]
\centering
\includegraphics[width=3.3in]{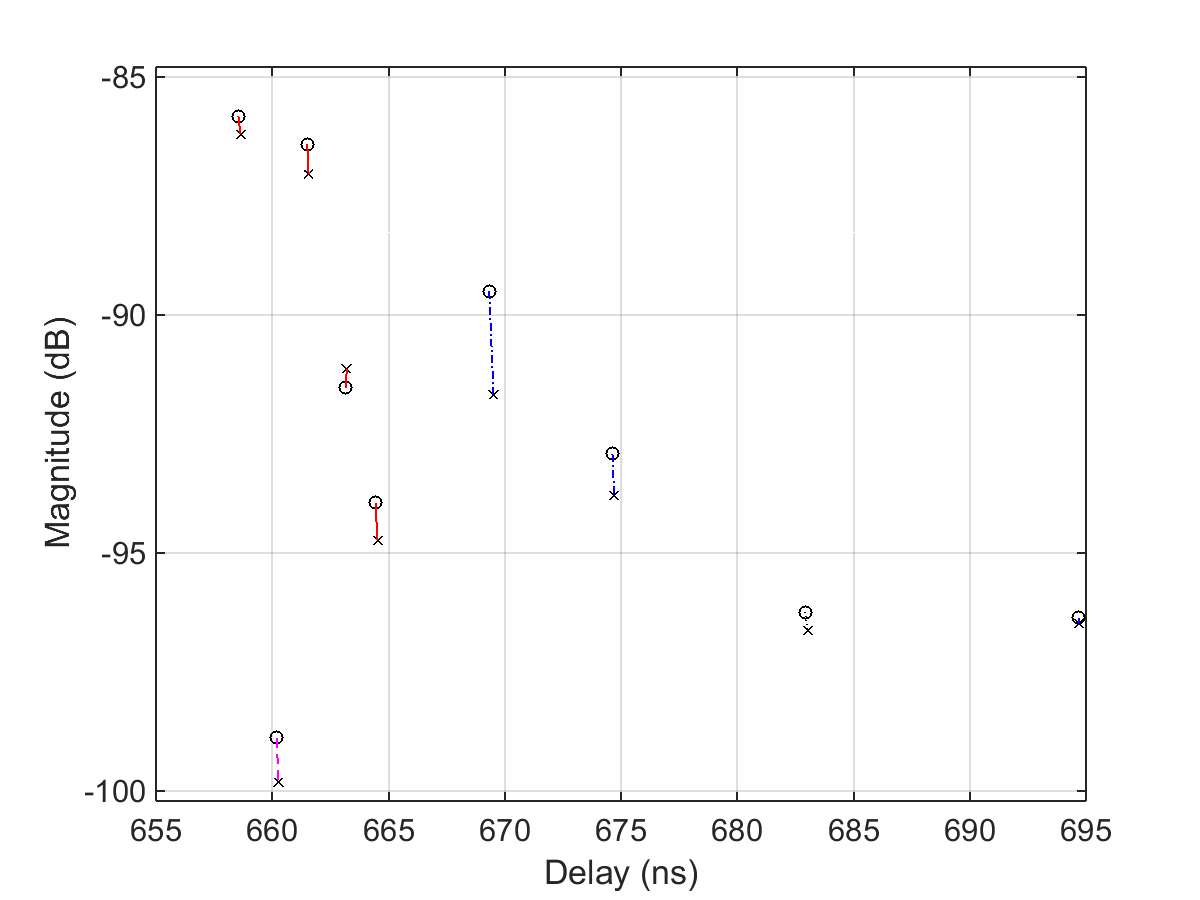}
\caption{Relating MPCs of current set (circles, 250 ms in Fig. \ref{ConnectTracks}) to MPCs of next set (crosses, 260 ms in Fig. \ref{ConnectTracks}).}
\label{ConnectTracksZoom}
\end{figure}

Our long-term tracking approach is similar to the tracking approaches described in \cite{he15} and \cite{czink06}. However, other than in these publications, our approach is not applied on direction-resolved measurement data, but calculates the multipath distance based on delay, power and Doppler frequencies estimates of individual MPCs. The main idea behind the algorithm is simple and can be best explained with Fig. \ref{ConnectTracksZoom}, where the circles indicate the averaged powers and delays of time-variant MPCs of the current set, and the crosses indicate the corresponding MPCs of the next recording set. As the channel is quite sparse, it is straightforward to relate the MPCs of different sets. The algorithm starts with the strongest MPC track in the current set, defines a two-dimensional search range and searches in the next set for possible candidates. In the next step of the algorithm, the delay change (Doppler frequency) of the current MPC track is used to predict the delay location of the MPC track in the next set. The same is done in the opposite direction; the delay change of the MPC track in the next set is used to predict the delay location of the MPC track in the current set. This additional prediction in the opposite direction increases the reliability of the tracking outcomes and is similar to the "two-way-matching" described in \cite{he15}. The deviation between the actual delay value and the predicted delay value is compared to different threshold values, again for both directions. Two MPC tracks are found to be related, if both deviations are below this threshold.

\begin{figure}[!t]
\centering
\includegraphics[width=3.3in]{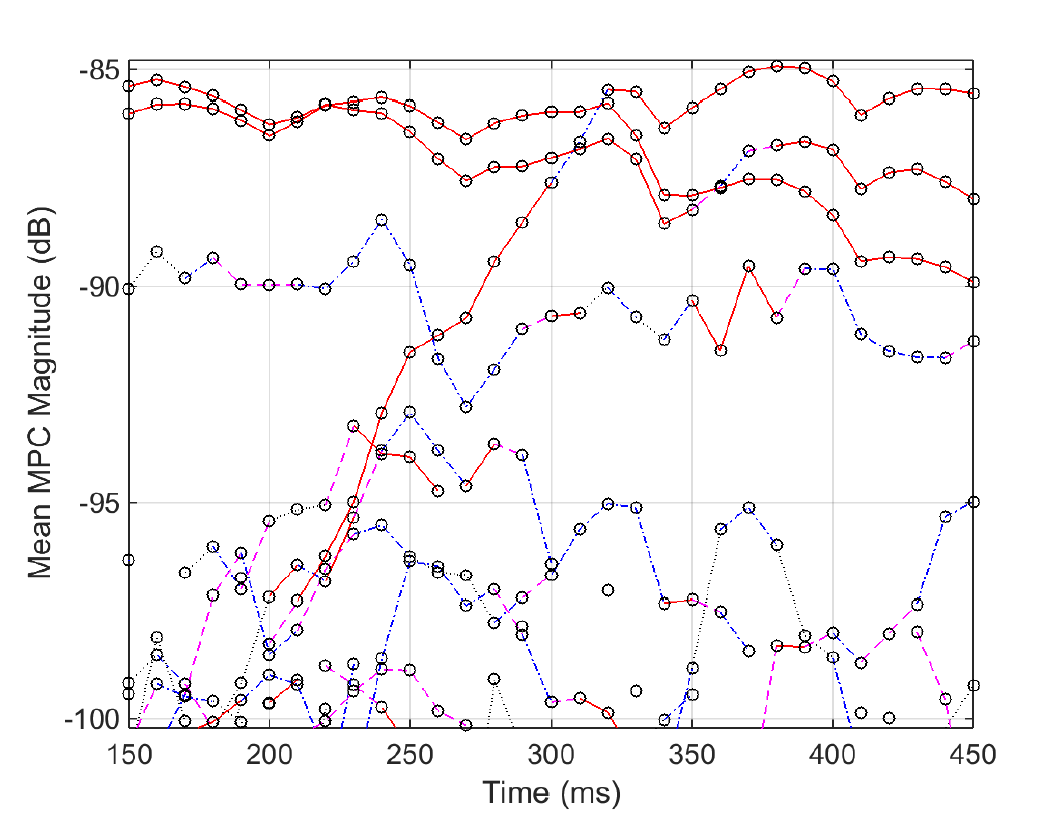}
\caption{Long-term tracking of MPC powers in the tunnel scenario. At the beginning two strong MPC tracks can be observed, whereas three dominant tracks can be found at the end of the diagram. The two-way delay prediction errors obtained from (\ref{gl:33}) and (\ref{gl:12}) are displayed as solid red line with errors $\chi \leq 0.1$ ns, dashed magenta $\chi \leq 0.2$ ns, dash-dotted blue line $\chi \leq 0.5$ ns and dotted black line $\chi \leq 1.0$ ns. }
\label{ConnectTracks}
\end{figure}

For the mathematical description of the long-term tracking algorithm, we have to note that $h(nT_r,uT_b) = h[n,u]$, where $T_r$ is the recording set period. We take the average MPC gain $\bar a_k$ and $\bar \tau_k$ from (\ref{gl:29}) and define the estimation parameters for the long-term tracking algorithm
\begin{equation}\label{gl:barwq}
\Theta_k = \{\bar a_k, \bar \tau_k, \psi_k, \nu_k \}
\end{equation}
with $k  \in \{1...K_i\}$ and $K_i$ being the number of tracks at  time instance $i$. Long-term time instances are denoted as $i$, in order to be clearly distinguishable from short-term time $n$ in Section II.C. We only consider full-lifetime MPCs $\psi_k \stackrel{!}{=} n_{snap}$, with $n_{snap}$ being the number of snapshots per set and define a subset
\begin{equation}\label{gl:barwq}
\boldsymbol{q}_k[i] = \left( \begin{array}{c}\bar a_k[i] \\ \bar \tau_k[i] \\ \nu_k[i]  
\end{array} \right).
\end{equation}
We start again with the strongest MPC $\hat q[i]$ in $i=i_k^{start}=1$ and search for candidates using 
\begin{equation}\label{gl:bla}
x_k^c=\begin{cases}1, & \text{if $d \left(\boldsymbol{\hat q}[i],\{ \boldsymbol{q}_k[i+1]\}_{k=1}^{K_{i+1}} \right) \leq \boldsymbol{\xi}_q$}\\
0 &\text{otherwise,} \end{cases}
\end{equation}
where d(.,.) is defined as in (\ref{gl:9}) and $\boldsymbol{\xi}_q$ is set empirically to a maximum delay change value of $\xi_\tau$ = 1 ns and a maximum magnitude change value $\xi_a$ = 5 dB. The identified candidates
\begin{equation}\label{gl:barwq}
 x_r^c \{ \boldsymbol{q}_c[i+1]\}_{c=1}^{C} = \{ \boldsymbol{q}^1_r[i+1],\boldsymbol{q}^2_r[i+1],...,\boldsymbol{q}^{C}_r[i+1]\}
\end{equation}
 are used together with the linear prediction $\tilde \tau_r[i+1]$ from model $G$ in
\begin{equation}\label{gl:33}
\tilde \tau_r[i+1] = G(\bar \tau_k[i], \nu_k[i]) = \bar \tau_k[i] - \nu_k[i] f_c T_r
\end{equation}
to find the closest candidate in terms of delay change prediction
\begin{equation}\label{gl:barwq}
\eta = \underset{c}{\operatorname{argmin}} ~d \left(\tilde \tau_r[i+1], x_r^c \{ \bar \tau_c [i+1] \}_{c=1}^C \right).
\end{equation}
Now, we use the selected candidate $\eta$ to predict the delay in the opposite direction
\begin{equation}\label{gl:35}
\tilde \tau_k[i] =\bar \tau_r^\eta[i+1] + \nu_r^\eta[i+1] f_c T_r
\end{equation}
and check the condition
\begin{equation}\label{gl:bla}
x_{rk}^i=\begin{cases}1, & \text{if $ d_r  \leq \chi ~ \wedge ~  d_k \leq \chi$ }\\
0 &\text{otherwise,} \end{cases}
\end{equation}
where 
\begin{equation}\label{gl:bla}
d_r= d \left(\tilde \tau_r[i+1],\tau^\eta_r[i+1] \right) 
\end{equation}
\begin{equation}\label{gl:bla}
d_k=  d \left(\tilde \tau_k[i],\tau_k[i] \right) 
\end{equation}
The final output of the long-term tracking is
\begin{equation}\label{gl:bla}
\boldsymbol{q}_r[i+1] = \boldsymbol{q}_k[i] x_{rk}^{i}.
\end{equation}
The different threshold values for $\chi$ are coded in Fig. \ref{ConnectTracksZoom} and Fig. \ref{ConnectTracks}, where the solid red line indicates a delay prediction error below 0.1 ns, the dashed magenta line an error below 0.2 ns, the dash-dotted blue line below 0.5 ns and the dotted black line below 1.0 ns. 


\section{Results}

\subsection{Long-Term MPC evolution}
The long-term MPC tracking result for a small part of the measurement run is shown in Fig. \ref{ConnectTracks}, where each time instance represents the mean MPC powers of a time-variant channel impulse response as shown Fig. \ref{MPCTracking}. Likewise, the MPCs in Fig. \ref{MPCTracking} can be related to the MPC powers at 250 ms and 260 ms in Fig. \ref{ConnectTracks}. As previously mentioned, only full-lifetime MPC tracks are considered for the long-term tracking algorithm. The example in Fig. \ref{ConnectTracks} shows that most of the full lifetime MPC tracks can be related to adjacent MPC tracks, yet with different levels of trustworthiness as indicated by the different colors and line styles. Strong MPC tracks can be connected with higher reliability (solid red line), whereas weaker MPC tracks can only be related with higher tolerances. For this measurement run, only 5\% of the power of all (full lifetime) MPCs could not be related to adjacent MPC tracks. The lowest delay prediction error below 0.1 ns holds still for 72\% of the power of all connected MPC tracks.

From Fig. \ref{ConnectTracks}, several observations regarding the long-term evolution of MPCs can be made. The strongest 2-3 MPC tracks show a parallel and wave-like power fluctuation behavior. Another MPC track at around -90 dB is comparatively constant and shows also a (reversed) wave-like power fluctuation after 400 ms. Starting at 200 ms, a MPC track at -97.5 dB gains power and becomes the strongest MPC at 320 ms. Several weaker MPC tracks appear, reach a small power level and disappear within a short period, for instance the track between 350 and 400 ms. Very rarely (full-lifetime) MPC tracks appear in one set only, i.e. circles without any connecting line.
\subsection{Total Tracking Power Loss}
Summing up the power losses of all processing steps, 2 dB power loss due to DMCs in the short-term MPC tracking, around 5\% power loss due to the neglect of non-full-lifetime MPCs and again 5\% power loss due to losses in the long-term tracking, the total tracking power loss is 2.4 dB. This means that the final long-term tracking results and the drawn statistical conclusions reflect around 57\% of the measured channel power. The remaining power can be assigned to diffuse or other non-trackable MPCs, which have a minor impact on fading effects compared to specular MPCs. We therefore consider the temporal behavior of MPCs in the measured propagation channel well represented with our tracking results. 
\subsection{Statistical Characterization}
\begin{figure}[!t]
\centering
\includegraphics[width=3.2in]{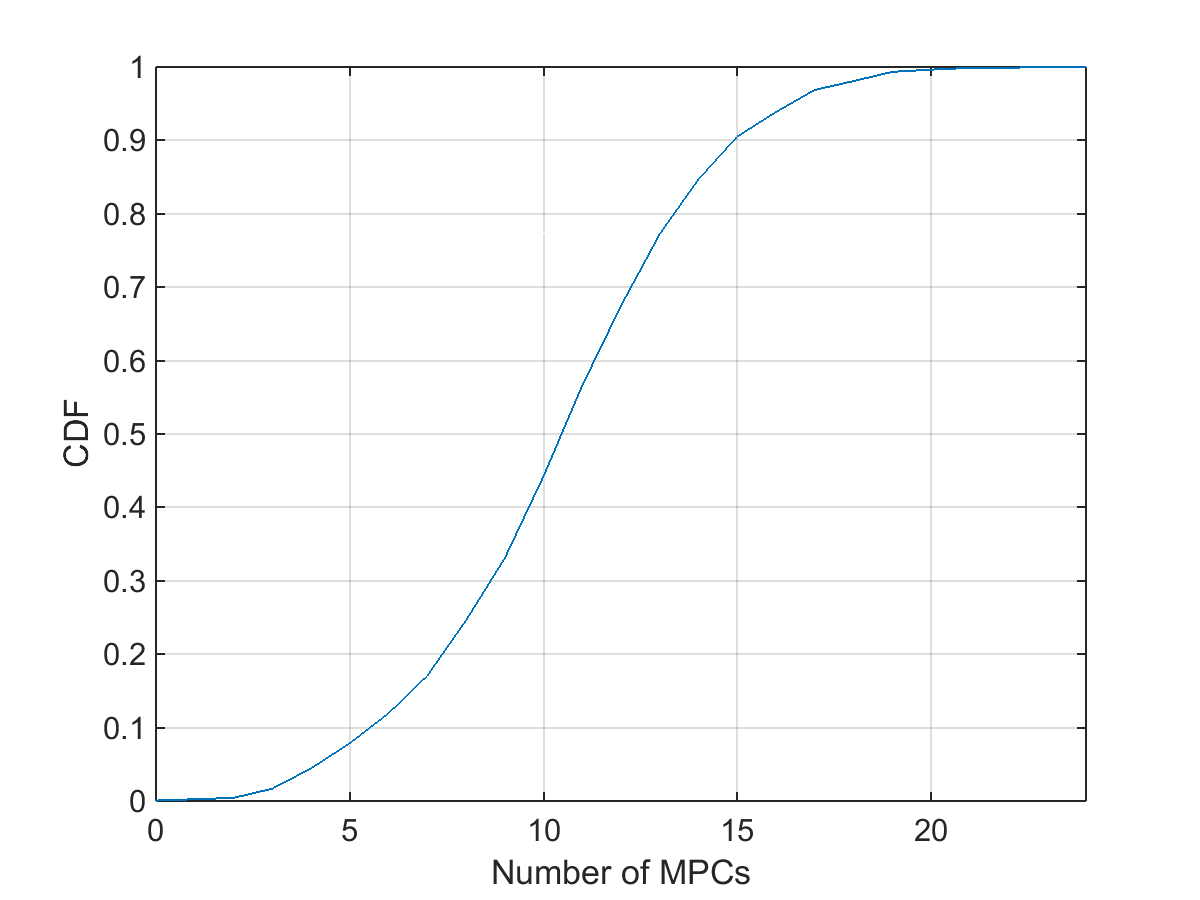}
\caption{Statistical distribution of number of MPCs in tunnel scenario.}
\label{NumTracksCDF}
\end{figure}

\begin{figure}[!t]
\centering
\includegraphics[width=3.2in]{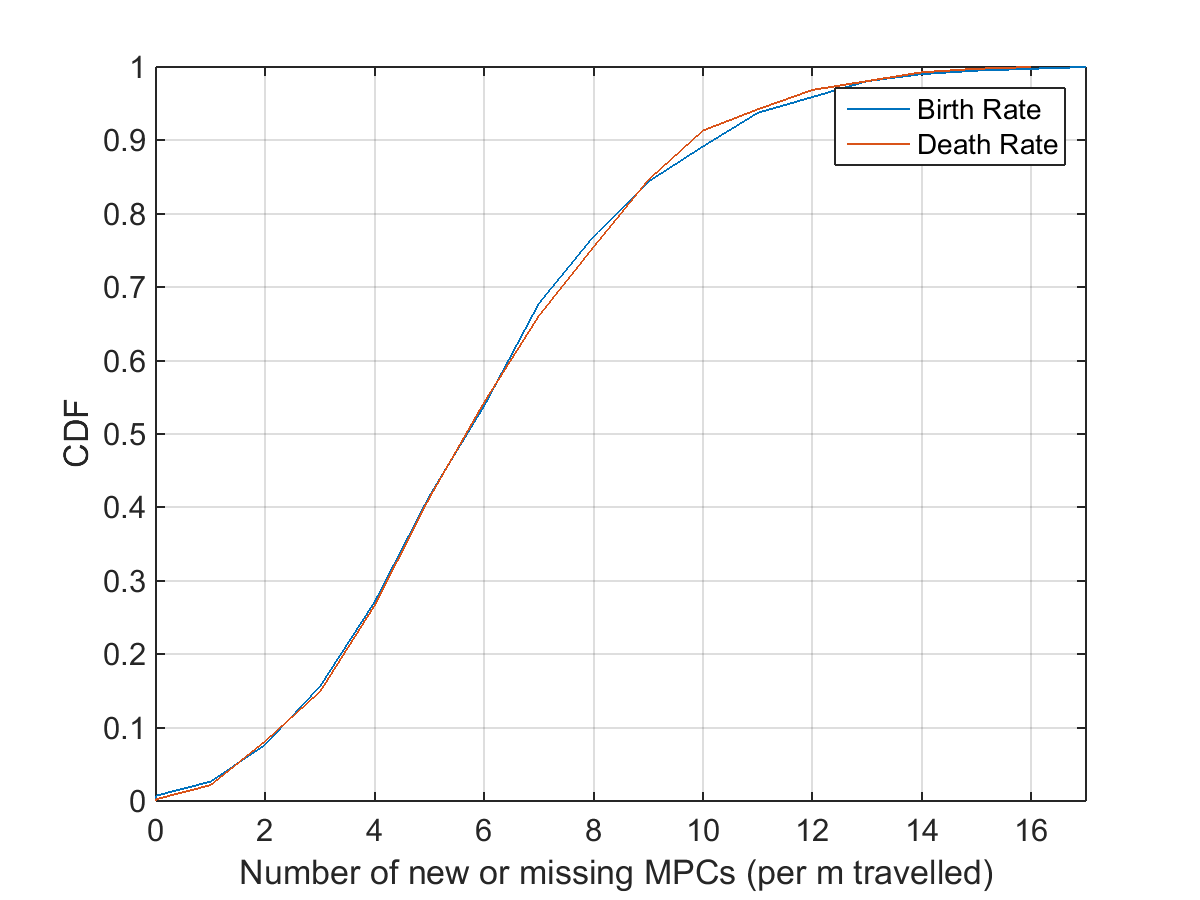}
\caption{Statistical distribution of birth/death rate of MPCs in tunnel scenario.}
\label{DeathBirthCDF}
\end{figure}

The wideband measurement data and the applied algorithms allow an extraction of all relevant MPC parameters. For this paper, we focus on the long-term MPC statistic results for the number of MPCs and the birth/death rate of MPCs. The tunnel measurement run with 1666 sets results in a distribution of the number of MPCs as depicted in Fig. \ref{NumTracksCDF}. The smooth shape of the CDF indicates a sufficiently large sample. Fig. \ref{DeathBirthCDF} shows the CDF of the birth and death rate of MPCs per meter travelled, i.e. the number of MPCs appearing or disappearing per cumulative distance of both vehicles.

\section{Summary}

Channel sounder measurements at 5.7 GHz with a bandwidth of 1 GHz lead to a sparse channel impulse response with favorable properties. The wide bandwidth makes highly accurate multipath components (MPC) estimation and tracking of individual MPCs feasible. Our proposed detection and tracking algorithms follow very well the progression of the channel gain fluctuations, with a power loss of only 2.5 dB. Compared to the GM-PHD filter, the proposed tracking algorithm has a comparable or better tracking performance and a significantly lower complexity. The proposed tracking method works flawless down to an SNR of 14 dB with a delay estimation error of 0.07 ns. In order to track the long-term evolution of MPC and relate tracks across measurement gaps of 6.4 ms, a long-term tracking method is applied. Based on these results, statistical distributions on the number of MPCs and the birth/death rate are drawn.

Apart from a better understanding of the time-variant propagation process, the proposed approach enables the development of more accurate channel models. The temporal evolution of highly-resolved individual MPCs can be related to physical objects and enhance the development of geometry-based channel models.



%


\bibliographystyle{IEEEtran}
\bibliography{literature}

\begin{thebibliography}{10}
\providecommand{\url}[1]{#1}
\csname url@samestyle\endcsname
\providecommand{\newblock}{\relax}
\providecommand{\bibinfo}[2]{#2}
\providecommand{\BIBentrySTDinterwordspacing}{\spaceskip=0pt\relax}
\providecommand{\BIBentryALTinterwordstretchfactor}{4}
\providecommand{\BIBentryALTinterwordspacing}{\spaceskip=\fontdimen2\font plus
\BIBentryALTinterwordstretchfactor\fontdimen3\font minus
  \fontdimen4\font\relax}
\providecommand{\BIBforeignlanguage}[2]{{%
\expandafter\ifx\csname l@#1\endcsname\relax
\typeout{** WARNING: IEEEtran.bst: No hyphenation pattern has been}%
\typeout{** loaded for the language `#1'. Using the pattern for}%
\typeout{** the default language instead.}%
\else
\language=\csname l@#1\endcsname
\fi
#2}}
\providecommand{\BIBdecl}{\relax}
\BIBdecl

\bibitem{renaudin08}
O.~Renaudin, V.~Kolmonen, P.~Vainikainen, and C.~Oestges, ``Wideband {MIMO}
  car-to-car radio channel measurements at 5.3 {GH}z,'' in \emph{Vehicular
  Technology Conference, 2008. VTC 2008-Fall. IEEE 68th}.\hskip 1em plus 0.5em
  minus 0.4em\relax IEEE, 2008, pp. 1--5.

\bibitem{karedal10}
J.~Karedal, F.~Tufvesson, T.~Abbas, O.~Klemp, A.~Paier, L.~Bernad{\'o}, and
  A.~F. Molisch, ``Radio channel measurements at street intersections for
  vehicle-to-vehicle safety applications,'' in \emph{Vehicular Technology
  Conference (VTC 2010-Spring), 2010 IEEE 71st}.\hskip 1em plus 0.5em minus
  0.4em\relax IEEE, 2010, pp. 1--5.

\bibitem{bernado14}
L.~Bernad{\'o}, T.~Zemen, F.~Tufvesson, A.~F. Molisch, and C.~F.
  Mecklenbrauker, ``Delay and {D}oppler spreads of nonstationary vehicular
  channels for safety-relevant scenarios,'' \emph{Vehicular Technology, IEEE
  Transactions on}, vol.~63, no.~1, pp. 82--93, 2014.

\bibitem{richter05}
A.~Richter, ``Estimation of radio channel parameters: Models and
  algorithms.''\hskip 1em plus 0.5em minus 0.4em\relax ISLE, 2005.

\bibitem{molisch05}
A.~F. Molisch, ``Ultrawideband propagation channels-theory, measurement, and
  modeling,'' \emph{Vehicular Technology, IEEE Transactions on}, vol.~54,
  no.~5, pp. 1528--1545, 2005.

\bibitem{win98}
M.~Z. Win and R.~A. Scholtz, ``On the energy capture of ultrawide bandwidth
  signals in dense multipath environments,'' \emph{IEEE Communications
  Letters}, vol.~2, no.~9, pp. 245--247, 1998.

\bibitem{lee09}
J.-Y. Lee, J.-H. Lee, S.-D. Kim, J.-H. Jeong, W.-H. Kim, G.-Y. Ha, C.-S. Jung,
  J.-E. Oh, D.-W. Ha, S.-J. Kong \emph{et~al.}, ``{UWB} propagation
  measurements in vehicular environments,'' in \emph{Radio and Wireless
  Symposium, 2009. RWS'09. IEEE}.\hskip 1em plus 0.5em minus 0.4em\relax IEEE,
  2009, pp. 236--239.

\bibitem{he15}
R.~He, O.~Renaudin, V.~M. Kolmonen, K.~Haneda, Z.~Zhong, B.~Ai, and C.~Oestges,
  ``A dynamic wideband directional channel model for vehicle-to-vehicle
  communications,'' \emph{IEEE Transactions on Industrial Electronics},
  vol.~62, no.~12, pp. 7870--7882, Dec 2015.

\bibitem{karedal09}
J.~Karedal, F.~Tufvesson, N.~Czink, A.~Paier, C.~Dumard, T.~Zemen, C.~F.
  Mecklenbrauker, and A.~F. Molisch, ``A geometry-based stochastic {MIMO} model
  for vehicle-to-vehicle communications,'' \emph{Wireless Communications, IEEE
  Transactions on}, vol.~8, no.~7, pp. 3646--3657, 2009.

\bibitem{czink06}
N.~Czink, C.~Mecklenbrauker, and G.~Del~Galdo, ``A novel automatic cluster
  tracking algorithm,'' in \emph{Personal, Indoor and Mobile Radio
  Communications, 2006 IEEE 17th International Symposium on}, Sept 2006, pp.
  1--5.

\bibitem{soh02}
M.~Steinbauer, H.~{\"O}zcelik, H.~Hofstetter, C.~Mecklenbr{\"a}uker, and
  E.~Bonek, ``How to {Q}uantify {M}ultipath {S}eparation,'' \emph{IEICE
  Transactions C: on Electronics, Special Issue on Signals, Systems, and
  Electronics Technology}, no.~3, pp. 552--557, Feb. 2002.

\bibitem{froehle12}
M.~Froehle, P.~Meissner, and K.~Witrisal, ``Tracking of {UWB} multipath
  components using probability hypothesis density filters,'' in
  \emph{Ultra-Wideband (ICUWB), 2012 IEEE International Conference on}, Sept
  2012, pp. 306--310.

\bibitem{paschalidis08}
P.~Paschalidis, M.~Wisotzki, A.~Kortke, W.~Keusgen, and M.~Peter, ``A wideband
  channel sounder for car-to-car radio channel measurements at 5.7 {GH}z and
  results for an urban scenario,'' in \emph{Vehicular Technology Conference,
  2008. VTC 2008-Fall. IEEE 68th}.\hskip 1em plus 0.5em minus 0.4em\relax IEEE,
  2008, pp. 1--5.

\bibitem{roy89}
R.~Roy and T.~Kailath, ``Esprit-estimation of signal parameters via rotational
  invariance techniques,'' \emph{Acoustics, Speech and Signal Processing, IEEE
  Transactions on}, vol.~37, no.~7, pp. 984--995, 1989.

\bibitem{schmidt86}
R.~O. Schmidt, ``Multiple emitter location and signal parameter estimation,''
  \emph{Antennas and Propagation, IEEE Transactions on}, vol.~34, no.~3, pp.
  276--280, 1986.

\bibitem{santos08}
T.~Santos, J.~Karedal, P.~Almers, F.~Tufvesson, and A.~F. Molisch, ``Scatterer
  detection by successive cancellation for {UWB}-method and experimental
  verification,'' in \emph{Vehicular Technology Conference, 2008. VTC Spring
  2008. IEEE}.\hskip 1em plus 0.5em minus 0.4em\relax IEEE, 2008, pp. 445--449.

\bibitem{Cramer02}
R.-M. Cramer, R.~Scholtz, and M.~Win, ``Evaluation of an ultra-wide-band
  propagation channel,'' \emph{Antennas and Propagation, IEEE Transactions on},
  vol.~50, no.~5, pp. 561--570, May 2002.

\bibitem{Haneda03}
K.~Haneda and J.-i. Takada, ``An application of {SAGE} algorithm for {UWB}
  propagation channel estimation,'' in \emph{Ultra Wideband Systems and
  Technologies, 2003 IEEE Conference on}, Nov 2003, pp. 483--487.

\bibitem{falsi06}
C.~Falsi, D.~Dardari, L.~Mucchi, and M.~Z. Win, ``Time of arrival estimation
  for {UWB} localizers in realistic environments,'' \emph{EURASIP Journal on
  Applied Signal Processing}, vol. 2006, pp. 152--152, 2006.

\bibitem{kaiser80}
J.~Kaiser and R.~W. Schafer, ``On the use of the {I0}-sinh window for spectrum
  analysis,'' \emph{Acoustics, Speech and Signal Processing, IEEE Transactions
  on}, vol.~28, no.~1, pp. 105--107, 1980.

\bibitem{vo06}
B.-N. Vo and W.-K. Ma, ``The {G}aussian mixture probability hypothesis density
  filter,'' \emph{Signal Processing, IEEE Transactions on}, vol.~54, no.~11,
  pp. 4091--4104, 2006.

\bibitem{salmi09}
J.~Salmi, A.~Richter, and V.~Koivunen, ``Detection and tracking of {MIMO}
  propagation path parameters using state-space approach,'' \emph{Signal
  Processing, IEEE Transactions on}, vol.~57, no.~4, pp. 1538--1550, 2009.

\bibitem{meifang}
M.~Zhu, J.~Vieira, Y.~Kuang, K.~Astrom, A.~Molisch, and F.~Tufvesson,
  ``Tracking and positioning using phase information from estimated multi-path
  components,'' in \emph{Communication Workshop (ICCW), 2015 IEEE International
  Conference on}, June 2015, pp. 712--717.

\bibitem{Fleury99}
B.~H. Fleury, M.~Tschudin, R.~Heddergott, D.~Dahlhaus, and K.~I. Pedersen,
  ``Channel parameter estimation in mobile radio environments using the sage
  algorithm,'' \emph{IEEE Journal on Selected Areas in Communications},
  vol.~17, no.~3, pp. 434--450, Mar 1999.

\bibitem{Ba07}
B.-T. Vo, B.-N. Vo, and A.~Cantoni, ``Analytic implementations of the
  cardinalized probability hypothesis density filter,'' \emph{Signal
  Processing, IEEE Transactions on}, vol.~55, no.~7, pp. 3553--3567, July 2007.

\end{thebibliography}

\begin{IEEEbiography}[{\includegraphics[width=1in,height=1.25in,clip,keepaspectratio]{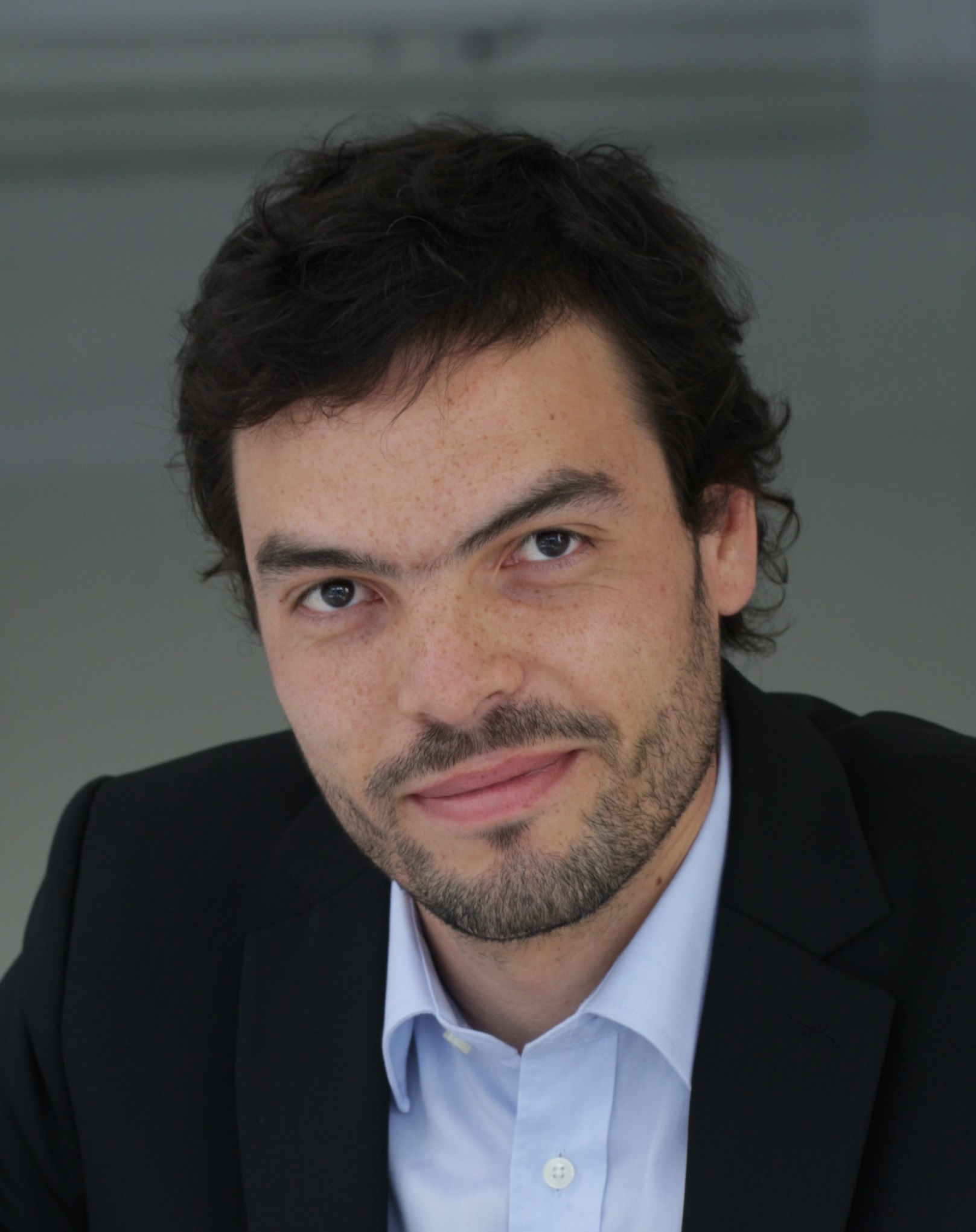}}]
{Kim Mahler} received his M.Sc. degree with honors from the EECS department from the Technical University of Berlin and an M.A. degree from the Berlin University of Arts / University of St. Gallen. Kim is with the Wireless Communications and Networks department at the Fraunhofer Heinrich Hertz Institute and working as a researcher in projects related to vehicular communications. His research interests involve extraction of time-variant wideband multipath components and parametrization of geometry-based stochastic channel models.
\end{IEEEbiography}
\vspace*{-3\baselineskip}
\begin{IEEEbiography}[{\includegraphics[width=1in,height=1.25in,clip,keepaspectratio]{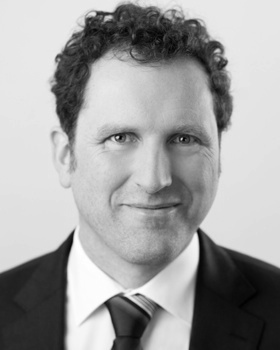}}]
{Wilhelm Keusgen} received the Dipl.-Ing. (M.S.E.E.) and Dr.-Ing. (Ph.D.E.E.) degrees from the RWTH Aachen University, Aachen, Germany, in 1999 and 2005, respectively. From 1999 to 2004, he was with the Institute of High Frequency Technology, RWTH Aachen University. Since 2004 he is heading a research group for mm-waves and advanced transceiver technologies at the Fraunhofer Heinrich Hertz Institute, located in Berlin, Germany. His main research areas are millimeter wave communications for 5G, measurement and modeling of wireless propagation channels, multiple antenna systems, and compensation of transceiver impairments. Since 2007 he also has a lectureship at the Technical University Berlin.
\end{IEEEbiography}
\vspace*{-3\baselineskip}
\begin{IEEEbiography}[{\includegraphics[width=1in,height=1.25in,clip,keepaspectratio]{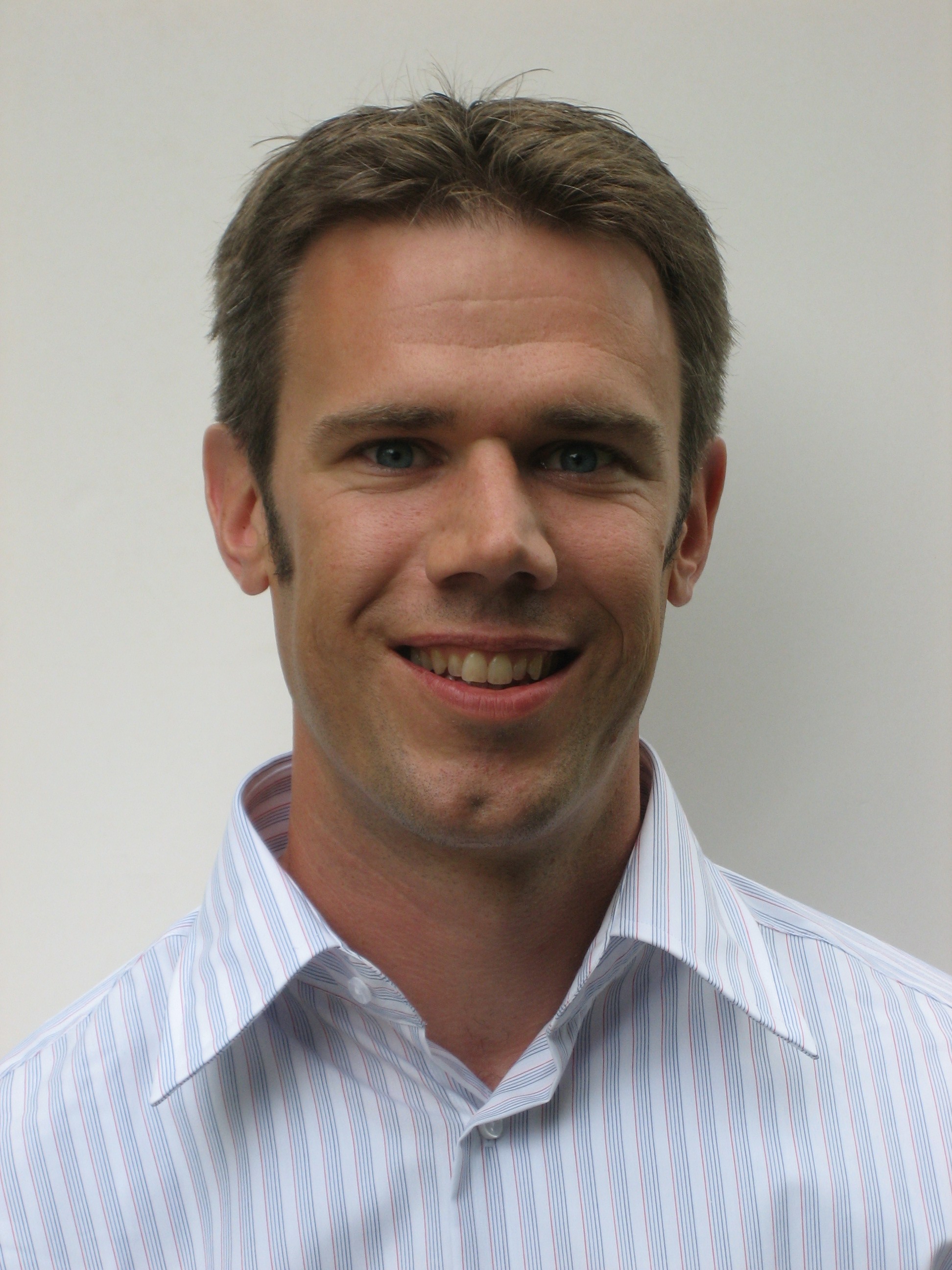}}]
{Fredrik Tufvesson} received his Ph.D. in 2000 from Lund University in Sweden. After two years at a startup company, he joined the department of Electrical and Information Technology at Lund University, where he is now professor of radio systems. His main research interests are channel modelling, measurements and characterization for wireless communication, with applications in various areas such as massive MIMO, UWB, mm wave communication, distributed antenna systems, radio based positioning and vehicular communication. Fredrik has authored around 60 journal papers and 120 conference papers, recently he got the Neal Shepherd Memorial Award for the best propagation paper in IEEE Transactions on Vehicular Technology.
\end{IEEEbiography}
\vspace*{-3\baselineskip}
\begin{IEEEbiography}[{\includegraphics[width=1in,height=1.25in,clip,keepaspectratio]{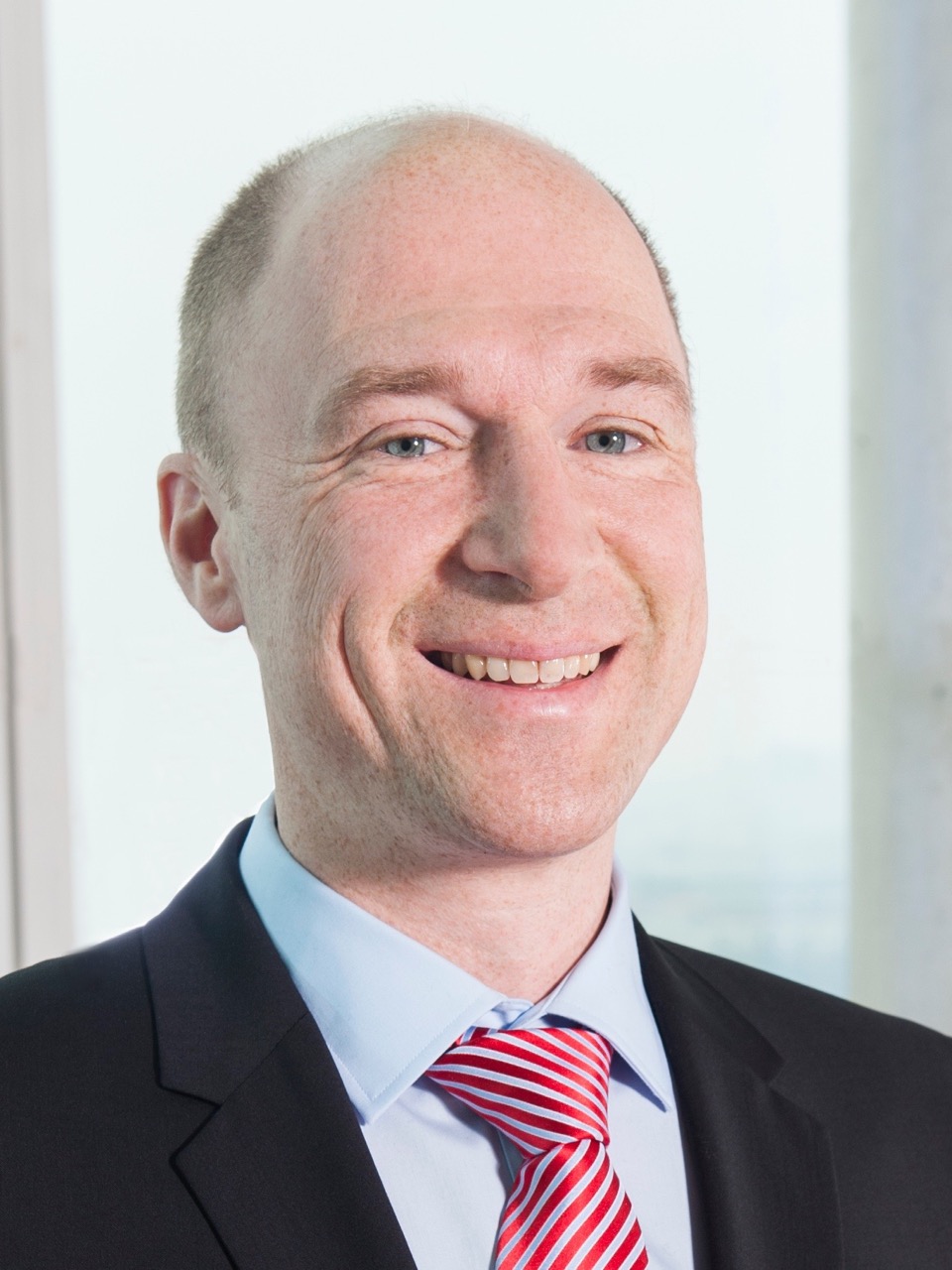}}]
{Thomas Zemen} (S'03--M'05--SM'10) received the Dipl.-Ing. degree in electrical engineering in 1998, the doctoral degree in 2004 and the Venia Docendi in 2013, all from Vienna University of Technology. He worked at Siemens Austria from 1998 to 2003; at FTW Telecommunications Research Center Vienna from 2003 to 2015, where he was Head of the "Signal and Information Processing" department since 2008. Since 2014 Thomas Zemen has been Senior Scientist at AIT Austrian Institute of Technology leading the research group for ultra-reliable wireless machine-to-machine communications. He is the author or coauthor of four books chapters, 32 journal papers and more than 80 conference communications. His research interests focus on reliable low-latency wireless communications, vehicular channel measurements and modelling, time-variant channel estimation, and cooperative communication systems. Dr. Zemen is an External Lecturer with the Vienna University of Technology and serves as Editor for the IEEE Transactions on Wireless Communications. 
\end{IEEEbiography}
\vspace*{-3\baselineskip}
\begin{IEEEbiography}[{\includegraphics[width=1in,height=1.25in,clip,keepaspectratio]{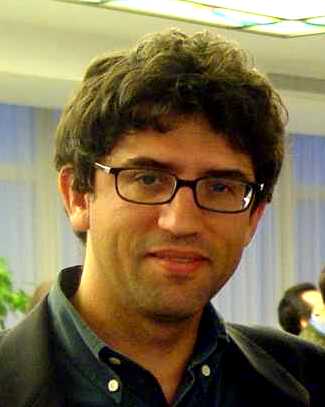}}]
{Giuseppe Caire} (S'92--M'94--SM'03--F'05) was born in Torino, Italy, in 1965. He received the B.Sc. in Electrical Engineering  from Politecnico di Torino (Italy), in 1990, the M.Sc. in Electrical Engineering from Princeton University in 1992 and the Ph.D. from Politecnico di Torino in 1994.
He has been a post-doctoral research fellow with the European Space Agency (ESTEC, Noordwijk, The Netherlands) in 1994-1995, Assistant Professor in Telecommunications at the Politecnico di Torino, Associate Professor at the University of Parma, Italy, Professor with the Department of Mobile Communications at the Eurecom Institute,  Sophia-Antipolis, France, and he is currently a professor of Electrical Engineering with the Viterbi School of Engineering, University of Southern California, Los Angeles and an Alexander von Humboldt Professor with the Electrical Engineering and Computer Science Department of the Technical University of Berlin, Germany.
\end{IEEEbiography}

%








\end{document}